\documentclass[11pt,a4paper]{article}
\usepackage{amsfonts}
\usepackage{amssymb}
\usepackage{jheppub}
\usepackage{ulem}
\usepackage{extarrows}
\usepackage{multirow}
\usepackage{float}

\usepackage{inputenc}
\usepackage{xspace}
\usepackage{url}

\newcommand{\bea}{\begin{eqnarray}}
\newcommand{\eea}{\end{eqnarray}}
\newcommand{\bean}{\begin{eqnarray*}}
\newcommand{\eean}{\end{eqnarray*}}
\newcommand{\nn}{\nonumber\\}
\newcommand{\Sl}{\sum\limits}

\newcommand{\red}{\color{red}}

\def\W #1{\widetilde{#1}}

\def\Label#1{\label{#1}%
  \smash{\hbox to0pt{\raise1ex\hbox{\tiny[#1]}\hss}}}
\def\Label#1{\label{#1}}
\renewcommand{\eqref}[1]{eq.~(\ref{#1})}

\newcommand{\figref}[1]{figure~\ref{#1}}
\newcommand{\tabref}[1]{table~\ref{#1}}
\newcommand{\secref}[1]{section~\ref{#1}}
\newcommand{\appref}[1]{appendix~\ref{#1}}

\renewcommand{\red}{\color{black}}

\preprint{LU-TP 15-17}

\newcommand{\unipd}{Dipartimento di Fisica ed Astronomia, Universit\`a di Padova, Via Marzolo 8, 35131 Padova, Italy}
\newcommand{\pdinfn}{INFN, Sezione di Padova, Via Marzolo 8, 35131 Padova, Italy}
\newcommand{\SISSA}{SISSA, Via Bonomea 265, 34136 Trieste, Italy}

\allowdisplaybreaks
\title{On single and double soft behaviors in NLSM }
\author[a,1]{Yi-Jian Du%
\note{On leave from Center for Field Theory and Particle Physics, Department of Physics, Fudan University.}}
\author[b,c,d]{Hui Luo}
\affiliation[a]{Department of Astronomy and Theoretical Physics, Lund University,
  S\"olvegatan 14A, 223\,62~Lund, Sweden}
\affiliation[b]{\unipd}
\affiliation[c]{\pdinfn}
\affiliation[d]{\SISSA}

\emailAdd{yijian.du@thep.lu.se}
\emailAdd{hui.luo@pd.infn.it}
\abstract{In this paper, we study the single and double soft behaviors of tree level off-shell currents and on-shell amplitudes in nonlinear sigma model {\red(NLSM)}.
  We first propose and prove the leading soft behavior of the tree level currents with a single soft particle.
In the on-shell limit, this single soft emission becomes the Adler's zero.
Then we establish the leading and subleading soft behaviors of tree level currents with two {\red adjacent} soft particles.
With a careful analysis of the on-shell limit, we obtain the double soft behaviors of on-shell amplitudes {\red where the two soft particles are adjacent to each other}. {\red By applying Kleiss-Kuijf (KK) relation, we further obtain the leading and subleading behaviors of amplitudes with two nonadjacent soft particles.}}
\begin{document}
\maketitle
\section{Introduction}
%

A lot of physics are based on principles of symmetry.
Once a global symmetry $G$ of a theory is spontaneously broken into its subgroup $H$, the massless Goldstone bosons are in one-to-one correspondence with the broken generators.
{\red The unbroken symmetry generators $T_i$ and the broken generators $X_a$ have schematic commutation relations as
\bea
[T, T]\sim T, \, [X, T]\sim X, \, [X,X]\sim T
\eea
}
This global symmetry $G$ can be realized by proper defined fields of the Goldstone bosons, and the nonlinear sigma model can be used to describe the behaviors of those Goldstone bosons \cite{Coleman:1969sm,Callan:1969sn}.
Because the scattering amplitudes calculated at any point of the vacuum moduli are identical, the vacuum structure after spontaneously global symmetry breaking can be understood from the scattering amplitude point of view \cite{ArkaniHamed:2008gz}.

Roughly speaking, one could try to identify states in the Hilbert spaces of two different vacua by a rotation $e^{i Q^\alpha \theta^\alpha}$ along the moduli space, where $\theta^\alpha$ is an expectation value of an Goldstone boson connected to an charge operator $Q^\alpha$.
The rotated vacuum is a coherent state of zero momentum Goldstone bosons.
However, actually the operator $Q^\alpha$ for broken symmetries does not exist due to the fact that the state created by it has an divergent norm with the volume of space.

A proper strategy is proposed to study the vacuum as following.
To reveal the structure of group $G$ from the behaviors of amplitudes with additional Goldstone bosons\footnote{In the following discussions, we will use ``pion" in stead of ``Goldstone boson" as a more concrete physical description in the NLSM.} of zero momentum, which reflect a different vacuum, one can regulate these constant scalar fields by tiny momenta and send them to zero eventually with a very careful analysis.
As already shown in \cite{ArkaniHamed:2008gz}, the physical states in one vacuum can be expanded around those in another vacuum as
\bea\label{vacuum-expand}
|\psi\rangle_{\theta}
=|\psi\rangle+|\psi^{(1)}\rangle+|\psi^{(2)}\rangle+\dots.
\eea
Here, $|\psi\rangle$ contains only hard pions while the variation $|\psi^{(n)}\rangle$ contains information of $n$ soft {\red pions} with momenta $\tau q_1, \tau q_2, \dots, \tau q_n$ in which $\tau$ as a tiny constant number measures how soft these momenta are.

For the first order variation including single soft pion, the emission will vanish at zero momentum known as "Adler zero" \cite{Adler:1964um,Susskind:1970gf}.

For the second order variation, {\red the state containing double soft pions is related to an $[X^\alpha, X^\beta]$ transformation of the amplitude with only hard pions, where the action of $[X^\alpha, X^\beta]$ on any vector index is
\bea
([X^\alpha, X^\beta]V)^\rho=V^\alpha\delta^{\beta\rho}-V^\beta\delta^{\rho\alpha},\nonumber
\eea
and the amount of rotation on a hard pion $i$ is given by
\bea
\theta_i={1\over 2}\,{(p_\alpha-p_\beta)\cdot p_i\over (p_\alpha+p_\beta)\cdot p_i}.\nonumber
\eea}
Only if {\red the second order variation} involves additional two soft pions together with a complementary rotation of the generators for broken symmetries on the hard particles \cite{ArkaniHamed:2008gz}, 
{\red
\bea
|\psi\rangle^{(2)}&=& \lim_{\tau\rightarrow 0} |\psi+\pi_\alpha(\tau p)+\pi_\beta(\tau q)\rangle\nonumber\\
&+&\sum_i {1\over 2} {(p-q)\cdot p_i\over (p+q)\cdot p_i}|\pi_{\sigma_1}\cdots ([X^\alpha, X^\beta]\pi)_{\sigma_i}\cdots\pi_{\sigma_n}\rangle
\eea
}
then the double-soft pion emission can present the $G$ invariance of the amplitude at this order.
How much the rotation is taken on the hard particles can be determined from double-soft limit of the amplitude \cite{ArkaniHamed:2008gz}.

The single and double soft behaviors of pions can also be studied from a generalized BCFW method and from current algebra point of view \cite{Kampf:2012fn,Kampf:2013vha}. The same leading order soft behaviors also applies to fermions in Akulov-Volkov theory and supergravity theories in both three and four dimensions \cite{Chen:2014xoa,Chen:2014cuc}.

Inspired by a recent study on the sub- and subsub-leading single soft behaviors of the tree-level gravity amplitudes \cite{Cachazo:2014fwa} \footnote{more works, please see \cite{Casali:2014xpa, Schwab:2014xua, Bern:2014oka, He:2014bga, Larkoski:2014hta, Cachazo:2014dia, Afkhami-Jeddi:2014fia, Adamo:2014yya, Geyer:2014lca, Schwab:2014fia, Bianchi:2014gla, Broedel:2014fsa, Bern:2014vva, White:2014qia, Zlotnikov:2014sva, Kalousios:2014uva, Naculich:2014naa, Du:2014eca, Campiglia:2014yka, Liu:2014vva, Luo:2014wea, Broedel:2014bza, Schwab:2014sla, Chen:2014xoa, Chen:2014cuc, Larkoski:2014bxa, Vera:2014tda}},
which gives an evidence to some new symmetry in quantum gravity S-matrix \cite{Strominger:2013jfa, He:2014laa, Kapec:2014opa}, Cachazo, He and Yuan have proposed
the new behaviors with two soft particles in Galileon, Dirac-Born-Infeld, Einstein-Maxwell-Scalar, non-linear sigma model and Yang-Mills-Scalar theories \cite{Cachazo:2015ksa}. Further studies in gauge theories, super-Yang-Mills theory and string theory have been made \cite{Volovich:2015yoa, Klose:2015xoa}.

Among all these progresses, the single and double-soft behaviors of amplitudes in nonlinear sigma model (NLSM) with $SU(N)\times SU(N)\rightarrow SU(N)$ are following.
\begin{itemize}
\item[] \emph{Single soft behavior}

When the momentum of a particle $i$ tends to zero, i.e., $k_i=\tau p$ $(\tau\to 0)$, the $\tau^0$ order of an $n$-point tree level partial amplitude $A(1,\dots,i,\dots,2n)$ in NLSM behaves as
\bea
A^{(0)}(1,\dots,\W i,\dots,2n)=0,~~\Label{Eq:On-shell-Leading-Single-Soft}
\eea
which is exactly the Adler's zero result \cite{Adler:1964um,Susskind:1970gf} and studied by a generalized BCFW recursion \cite{Kampf:2012fn,Kampf:2013vha}.
\item[] \emph{Double soft behavior}

When two adjacent particles $i$, $i+1$ are soft, i.e., $k_i=\tau p$ and $k_{i+1}=\tau q$ with $\tau\to 0$, the partial amplitude $A(1,\dots,\W i,\W {i+1}, \dots, 2n)$ behaves as
\bea
A(1,\dots,\W i,{\red\W {i+1}}, \dots, 2n)=\left(\tau^0 \mathbb{S}_{i,i+1}^{(0)}+\tau^1 \mathbb{S}_{i,i+1}^{(1)}\right)A(1,\dots,i-1,i+2,\dots,2n)+\mathcal{O}(\tau^2).\Label{Eq:On-shell-Double-Soft}
\eea
In the above expression, the leading double soft factor $\mathbb{S}_{i,i+1}^{(0)}$ is
\bea
\mathbb{S}_{i,i+1}^{(0)}&=&\left(-{1\over 2F^2}\right){1\over 2}\left[{k_{i-1}\cdot(p-q)\over k_{i-1}\cdot(p+q)}+{k_{i+2}\cdot(q-p)\over k_{i+2}\cdot(q+p)}\right]\Label{Eq:On-shell-Leading-Double-Soft}
\eea
which is related to the vacuum structure as discussed in \cite{ArkaniHamed:2008gz}.
The subleading double-soft factor $\mathbb{S}_{i,i+1}^{(1)}$ is
\bea
\mathbb{S}_{i,i+1}^{(1)}&=&\left(-{1\over 2F^2}\right)(p\cdot q)\left[{k_{i-1}\cdot q\over (k_{i-1}\cdot(p+q))^2}+{k_{i+2}\cdot p\over (k_{i+2}\cdot(p+q))^2}\right]\Label{Eq:On-shell-Subleading-Double-Soft}\nn
&&~~~~+\left(-{1\over 2F^2}\right)\left[{p_{\mu}q_{\nu}\over k_{i-1}\cdot (p+q)}{\cal J}_{i-1}^{\mu\nu}+{q_{\mu}p_{\nu}\over k_{i+2}\cdot (p+q)}{\cal J}_{i+2}^{\mu\nu}\right],
\eea
in which the angular momentum operator ${\cal J}_a^{\mu\nu}$ of a scalar particle $a$ is defined by
\bea
{\cal J}_a^{\mu\nu}\equiv k_a^{\mu}{\partial\over \partial k_{a,\nu}}-k_a^{\nu}{\partial\over \partial k_{a,\mu}}.~~\Label{Eq:AngularMomentum}
\eea
Both the Adler's zero \eqref{Eq:On-shell-Leading-Single-Soft} and the leading order of the  double soft behavior \eqref{Eq:On-shell-Leading-Double-Soft} in NLSM were studied by the new BCFW recursion  \cite{Kampf:2013vha} and by the Cachazo-He-Yuan (CHY) \cite{Cachazo:2013hca, Cachazo:2013iea,Cachazo:2014xea} formula \cite{Cachazo:2015ksa}. Moreover, the subleading order \eqref{Eq:On-shell-Subleading-Double-Soft} of the double soft behavior was also proposed in \cite{Cachazo:2015ksa}.
\end{itemize}
Although the single and double soft behaviors of the tree-level on-shell amplitudes in NLSM have already been investigated from different compact amplitude constructions, i.e., BCFW recursion and CHY formula, no insight into Feynman diagrams has been provided and the corresponding soft behaviors of off-shell currents have not been fully understood.
 Berends-Giele recursion relation \cite{Berends:1987me} as a recursive construction of amplitudes from Feynman diagrams provides a way to systematically study the (single and double) soft behaviors. By this means, we can find out how the soft behaviors emerge in their diagram structures.

This work is devoted to understanding the single and double soft behaviors of the off-shell currents in NLSM from the Berends-Giele recursion relation ({{\red A brief review of NLSM, Feynamn rules and Berends-Giele recursion are given in \appref{app:Feynman-rules-BG}}}).
With the Cayley parameterization in NLSM, we find that the tree-level current $J(2,\dots,\W i,\dots,2n)$\footnote{\red In this paper, the particle $1$ is chosen as the off-shell leg.}  ({\red defined by \eqref{B-G}}) with $i$ as the soft particle behaves as
\bea
J(2,\dots,i-1,\W i,i+1,\dots,2n)=\left\{
                          \begin{array}{cc}
                            0 & (\text{$i$ is even}) \\
                           \left(1\over 2F^2\right)J(2,\dots,i-1)J(i+2,\dots,2n)  &  (\text{$i$ is odd}) \\
                          \end{array}\right\}+\mathcal{O}(\tau),\Label{Eq:Off-shell-Leading-Single-Soft}
\eea
where $k_i=\tau p$ and $\tau\to 0$.
In the on-shell limit, the amplitude $A(1,2,\dots,i,\dots,2n)$ is presented by $\lim\limits_{P_{2,2n}^2\to 0}(-1)P_{2,2n}^2J(2,\dots,i,\dots,2n)$ (where $P_{2,2n}$ denotes the sum of momenta of all on-shell particles $2$, ..., $2n$) and achieves Adler's zero \eqref{Eq:On-shell-Leading-Single-Soft} for both even and odd $i$ when $k_i\to 0$.

We then study the leading and sub-leading order behaviors of the off-shell currents with two adjacent soft particles. According to different possible positions of the soft particles, we consider the following two cases:
 \begin{itemize}
 \item[] {(A)} the current with one soft particle adjacent to the off-shell particle,
 \item[] {(B)} the current with no soft particle adjacent to the off-shell particle.
 \end{itemize}
In both types, the momentum of the off-shell line is the sum of the momenta of all the other on-shell particles.
{\red Recalling that the momentum conservation imposes a constraint on external momenta, we can arbitrarily choose the dependent momentum and the expression \eqref{Eq:On-shell-Double-Soft} for on-shell amplitudes indeed holds for all choices.}
The above two cases {\red (A) and (B) (for off-shell currents)} actually respect to the different choices of the dependent momentum for the corresponding on-shell amplitudes with two soft particles. Particularly, the first case corresponds to
choosing the momentum of a particle adjacent to one soft particle as the dependent one, while the second case corresponds to
choosing the momentum of a particle nonadjacent to any soft particle as the dependent one.
The double soft behaviors of these two types are following.
\begin{itemize}
\item For the first type, the current $J(\W 2,\W 3, 4, \dots,2n)$ with two soft particles $2$ and $3$ behaves as
\bea
J(\W 2,\W 3, 4,\dots, 2n)=\tau^0S_{2,3}^{(0)}J(4,\dots,2n)+\tau^1S_{2,3}^{(1)}J(4,\dots,2n)+\mathcal{O}(\tau),\Label{Off-shell-Leading-Subleading-Double-Soft0}
\eea
where $k_2=\tau p$ and $k_3=\tau q$. The leading and subleading double soft factors $S_{2,3}^{(0)}$ and $S_{2,3}^{(1)}$ are defined by
\bea
S_{2,3}^{(0)}=\left({1\over 2F^2}\right){k_{4}\cdot p\over k_{4}\cdot(q+p)},~~~~
S_{2,3}^{(1)}=\left(-{1\over 2F^2}\right)\left[(p\cdot q){k_4\cdot p\over (k_4\cdot(p+q))^2}+{q_{\mu}p_{\nu}{\cal J}_4^{\mu\nu}\over k_4\cdot(p+q)}\right].
\eea
These factors apparently are different from the factors for on-shell amplitudes \eqref{Eq:On-shell-Leading-Double-Soft} and \eqref{Eq:On-shell-Subleading-Double-Soft}. However, by taking the on-shell limit very carefully, we return to the on-shell double soft behavior \eqref{Eq:On-shell-Double-Soft} with the factors \eqref{Eq:On-shell-Leading-Double-Soft} and \eqref{Eq:On-shell-Subleading-Double-Soft}.
The behavior of current $J(2,\dots,\W{2n-1},\W{2n})$ with $2n-1$, $2n$ as the soft particles can be obtained by considering the reflection symmetry directly.

\item For the second type, we propose that the leading and subleading orders of the current $J(2,\dots,i-1,\W i,\W {i+1},i+2,\dots,n)$ with two soft particles $i$, $i+1$ are given by
\bea
&&J(2,\dots,i-1,\W i,\W {i+1},i+2,\dots,2n)~~~~~~~\Label{Off-shell-Leading-Subleading-Double-Soft1}\\
&=&\tau^0 S_{i,i+1}^{(0)}J(2,\dots,i-1,i+2,\dots,2n)+\tau^1\Biggl[ S_{i,i+1}^{(1)}J(2,\dots,i-1,i+2,\dots,2n)\nn
&&~~~~~~~~~~~~~~~~~~~~~~~~~~+\left\{
                                           \begin{array}{cc}
                                            \left(1\over 2F^2\right) J^{(1)}(2,\dots,i-1,\W i)J(i+2,\dots,2n) &(\text{$i$ is even })  \\
                                            \left(1\over 2F^2\right) J(2,\dots,i-1)J^{(1)}(\W{i+1},i+2,\dots,2n)  &(\text{$i$ is odd}) \\
                                           \end{array}
                               \right\}\Biggr]+\mathcal{O}(\tau^2),\nonumber
\eea
where $k_i=\tau p$ and $k_{i+1}=\tau q$. The soft factors $S_{i,i+1}^{(0)}$ and $S_{i,i+1}^{(1)}$ are same with the on-shell double soft factors \eqref{Eq:On-shell-Leading-Double-Soft} and \eqref{Eq:On-shell-Subleading-Double-Soft}, i.e.,
\bea
S_{i,i+1}^{(0)}=\mathbb{S}_{i,i+1}^{(0)},~~~~S_{i,i+1}^{(1)}=\mathbb{S}_{i,i+1}^{(1)}.
\eea
 Again, the soft behavior at off-shell level in general is not same with the on-shell case \eqref{Eq:On-shell-Double-Soft}. There is an extra term expressed by the product of a subcurrent without any soft particle and the $\tau^1$ coefficient of a subcurrent with only one soft particle. However, by taking the double soft limit of the on-shell amplitude $-\lim\limits_{P_{2,2n}^2\to 0}P_{2,2n}^2J(2,\dots,i,i+1,\dots,2n)$, one arrives the on-shell double soft behavior \eqref{Eq:On-shell-Double-Soft}.

\end{itemize}

{\red Having the double soft behavior \eqref{Eq:On-shell-Double-Soft} of amplitudes with two adjacent soft particles, we will prove all the leading and subleading behaviors of amplitudes with two nonadjacent soft particles by Kleiss-Kuijf (KK) relation \cite{Kleiss:1988ne} in nonlinear sigma model (see the proof \cite{Chen:2013fya,Chen:2014dfa}).}

%
%
%
%

This paper is organized as follows. In \secref{sec:offshell-single-soft}, we discuss the single soft behavior of off-shell currents in NLSM.
With the help of the single soft results in \secref{sec:offshell-single-soft}, {\red we further study the leading and subleading behaviors of the off-shell currents with two adjacent soft particles in \secref{sec:offshell-double-soft}.}
Then we contract back the on-shell condition of the currents, and find out the consistent on-shell amplitudes results {\red \eqref{Eq:On-shell-Leading-Single-Soft} and \eqref{Eq:On-shell-Double-Soft}} in \secref{sec:on-shell-limit}. {\red Using KK relation, we derive all the double soft behaviors of amplitudes with two nonadjacent soft particles in \secref{sec:NonAdjSoft}.}
Finally, we give a conclusion and further discussion in the \secref{sec:conclusion}.
The Feynman rules and Berends-Giele recursion are given in \appref{app:Feynman-rules-BG}.
The computation details of the double-soft behaviors with the soft particles non-adjacent to the off-shell line are collected in \appref{app:appendix-nonadj}.

\section{Single soft behavior of currents}\label{sec:offshell-single-soft}
In this section, we study the single soft behavior of off-shell currents from the Berends-Giele recursion \eqref{B-G} of $U(N)$ nonlinear sigma model. The on-shell amplitudes of $SU(N)$ nonlinear sigma model can be obtained by taking on-shell limits \cite{Kampf:2012fn, Kampf:2013vha}. We first display explicit calculations on the  four- and six-point currents with one soft particle,  then provide a general proof of the leading order single soft behavior \eqref{Eq:Off-shell-Leading-Single-Soft}.

\subsection{Examples of lower-point currents}\label{single-example}
We write out expressions of the  four- and six-point currents with a single soft particle where the general single soft behavior
can be detected.

\subsubsection{Four-point current with one soft particle}
In the four-point current $J(2,3,4)$, any one of $2$, $3$ and $4$ can be chosen as a soft particle. Because of the reflection symmetry $J(2,3,4)=J(4,3,2)$, we only consider the cases in which $2$ or $3$ (equivalently, $3$ or $4$) carries a momentum $\tau p$ ($\tau\to 0$) and obtain
\bea
J(\W 2,3,4)=\left(-{1\over 2F^2}\right){i\over P_{2,4}^2}i(\tau p+k_4)^2=0+\mathcal{O}(\tau),\Label{Eq:Off-shell-Leading-Single-Soft-4pt1}
\eea
\bea
J(2,\W 3, 4)=\left(-{1\over 2F^2}\right){i\over P_{2,4}^2}i(k_2+k_4)^2={1\over 2F^2}+\mathcal{O}(\tau),\Label{Eq:Off-shell-Leading-Single-Soft-4pt2}
\eea
where $P_{i,j}(i<j)$ denotes the sum of momenta $\Sl_{r=i}^jk_r$.
Since $J(2)=J(4)=1$, the single soft behaviors represented by \eqref{Eq:Off-shell-Leading-Single-Soft-4pt1} and \eqref{Eq:Off-shell-Leading-Single-Soft-4pt2} agree with the general formula \eqref{Eq:Off-shell-Leading-Single-Soft} and  give rise the expected Adler's zero under the on-shell limit of the off-shell line.

\subsubsection{Six-point current with one soft particle}
%
\begin{figure}[H]
  \centering
 \includegraphics[width=1\textwidth]{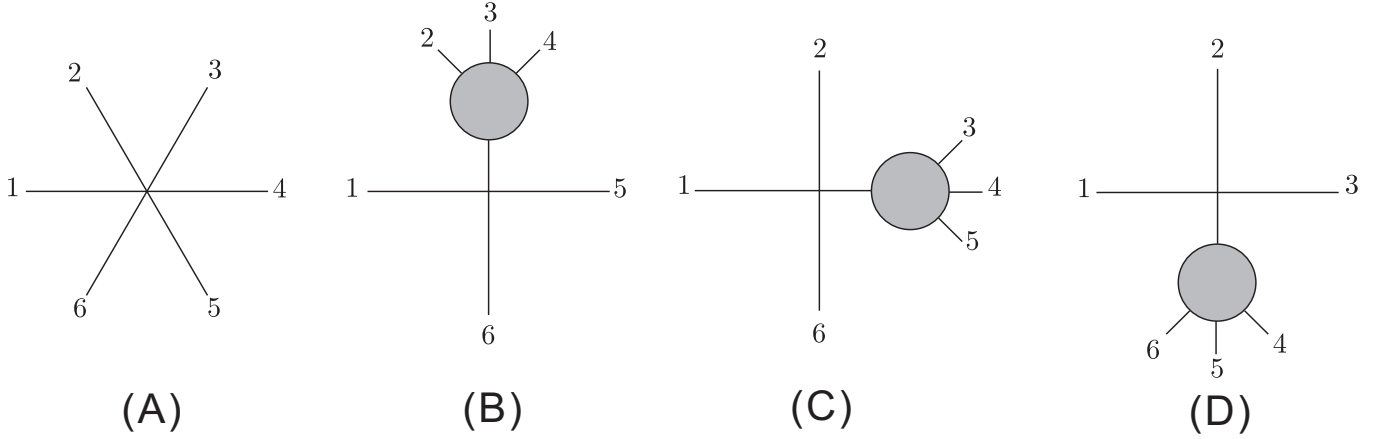}
 \caption{Diagrams which contribute to the six-point current $J(2,3,4,5,6)$. } \label{FigAll6PtDiagrams}
\end{figure}
The six-point current $J(2,3,4,5,6)$ is expressed in terms of four diagrams in \figref{FigAll6PtDiagrams} via the Berends-Giele recursion \eqref{B-G}. We now consider $2$, $3$ and $4$ as soft particles in turn. The case with $6$ or $5$ as soft particle can be obtained from the case with $2$ or $3$ as soft particle by reversing the order of all on-shell particles.

\subsubsection*{The particle $2$ is soft}
This is a boundary case where the soft particle $2$ ($k_2=\tau p$, $\tau\rightarrow 0$) is adjacent to the off-shell leg.

\begin{table}[H]{\centering
\begin{tabular}{c|c|c}
\hline
  Diagram &  $\tau^0$ contribution & LO results\\
  \hline \hline
  Diagram (B) & $-\left({1\over 2F^2}\right)(k_3+k_4+k_6)^2J(\W 2,3,4)$& 0 \\
  \hline
  Diagram (C) & $-\left({1\over 2F^2}\right)k_6^2J(3,4,5)$ & 0 \\
  \hline
  Diagram (A) & $+\left({1\over 2F^2}\right)^2\left(k_4+k_6\right)^2$& \multirow{2}{*}{0}\\
  \cline{1-2}
   Diagram (D)  & $-\left({1\over 2F^2}\right)(k_4+k_5+k_6)^2J(4,5,6)$ \\
  \hline
\end{tabular}
\caption{The $\tau^0$ order terms of the diagrams (A), (B), (C), (D) in  \figref{FigAll6PtDiagrams} with particle $2$ soft.}\Label{6p-single2-table1}}
\end{table}

As shown in the last column ``LO results'' (short for ``leading order results") in  \tabref{6p-single2-table1}, diagram (B) in \figref{FigAll6PtDiagrams} should vanish due to \eqref{Eq:Off-shell-Leading-Single-Soft-4pt1},
diagram (C) vanishes due to the on-shell condition $k_6^2=0$, while
diagrams (A) and (D) cancel with each other because of the explicit expression  $J(4,5,6)=\left(-{1\over 2F^2}\right){i\over P_{4,6}^2}i(k_4+k_6)^2$ in diagram (D).
%
%
Retrieving the propagator of the off-shell leg, i.e., $J(\W 2,3,4,5,6)=\tau^0{i\over P_{2,6}^2(k_2=0)}i  \left({\rm (A)+(B)+(C)+(D)}\right)$, we achieve the expected soft behavior \eqref{Eq:Off-shell-Leading-Single-Soft} with the soft particle index of an even number.

\subsubsection*{The particle $3$ is soft}
While the particle $3$ is soft, i.e., $k_3=\tau q$ with $\tau\to 0$, $J(2,\W 3,4,5,6)$ gains contributions as shown in  \tabref{6p-single3-table1} from diagrams in \figref{FigAll6PtDiagrams}.

\begin{table}[H]{\centering
\begin{tabular}{c|c|c}
\hline
  Diagram &  $\tau^0$ contribution & LO results\\
  \hline \hline
  Diagram (A) & $\left({1\over 2F^2}\right)^2\left(k_2+k_4+k_6\right)^2$& \multirow{2}{*}{0}\\
  \cline{1-2}
   Diagram (B)  & $-\left({1\over 2F^2}\right)(k_2+k_4+k_6)^2J(2,\W 3,4)$ \\
  \hline
    Diagram (C) & $-\left({1\over 2F^2}\right)\left(k_2+k_6\right)^2J(\W 3,4,5)$& 0 \\
  \hline
  Diagram (D) & $-\left({1\over 2F^2}\right)(k_2+k_4+k_5+k_6)^2J(2)J(4,5,6)$ & $-\left({1\over 2F^2}\right)(k_2+k_4+k_5+k_6)^2J(2)J(4,5,6)$\\
  \hline
\end{tabular}
\caption{The $\tau^0$ order terms of the diagrams (A), (B), (C), (D) in \figref{FigAll6PtDiagrams} with particle $3$ soft.}\Label{6p-single3-table1}}
\end{table}

In the third column ``LO results'' in table \ref{6p-single3-table1}, we find that diagrams (A) and (B) cancel with each other by writing explicitly subcurrent $J(2,\W 3,4)$ in diagram (B) with \eqref{Eq:Off-shell-Leading-Single-Soft-4pt2}.
Diagram (C) in figure \ref{FigAll6PtDiagrams} vanishes due to  \eqref{Eq:Off-shell-Leading-Single-Soft-4pt1}.
Contracting back the propagator of the off-shell line with the remaining non-zero contribution from diagram (D) and considering $P_{2,6}(k_3=0)=(k_2+k_4+k_5+k_6)^2$, we have
\bea
J(2,\W 3,4,5,6)=\left({1\over 2F^2}\right)^2J(2)J(4,5,6),
\eea
which agrees with the expected result \eqref{Eq:Off-shell-Leading-Single-Soft} with the odd $i$.

\subsubsection*{The particle $4$ is soft}
If the soft particle is $4$, we have $k_4=\tau p$ with $\tau\rightarrow 0$.

\begin{table}[H]{\centering
\begin{tabular}{c|c|c}
\hline
  Diagram &  $\tau^0$ contribution & LO results\\
  \hline \hline
  Diagram (A) & $\left({1\over 2F^2}\right)^2\left(k_2+k_6\right)^2$& \multirow{2}{*}{0}\\
  \cline{1-2}
   Diagram (C)  & $-\left({1\over 2F^2}\right)(k_2+k_6)^2J(3,\W 4,5)$ \\
  \hline
    Diagram (B) & $-\left({1\over 2F^2}\right)\left(k_2+k_3+k_6\right)^2J(2,3,\W4)$& 0 \\
  \hline
  Diagram (D) & $-\left({1\over 2F^2}\right)(k_2+k_5+k_6)^2J(\W 4,5,6)$ & 0\\
  \hline
\end{tabular}
\caption{The $\tau^0$ order terms of the diagrams (A), (B), (C), (D) in \figref{FigAll6PtDiagrams} with particle $4$ soft.}\Label{6p-single4-table1}}
\end{table}

In the last column "LO results" in \tabref{6p-single4-table1},
diagrams (A) and (C) cancel out when we apply the soft behavior of four-point current \eqref{Eq:Off-shell-Leading-Single-Soft-4pt2} to $J(3,\W 4,5)$ in diagram (C).
Diagrams (B) and (D) vanish because both $J(2,3,\W 4)$ and $J(\W 4,5,6)$, where the soft particle $4$ plays as an even number one, have to be zero due to \eqref{Eq:Off-shell-Leading-Single-Soft-4pt1}.
Contracting the propagator back, we get $
J(2,3,\W 4,5,6)=\tau^0\,0+\mathcal{O}(\tau)
$ which agrees with the single soft behavior  \eqref{Eq:Off-shell-Leading-Single-Soft} with even $i$.

\subsection{General proof}
The six-point example can be easily extended to a higher-point study. Now let us elaborate a general proof for the single soft behavior \eqref{Eq:Off-shell-Leading-Single-Soft}. According to six-point examples, we will consider the following three cases as well
\begin{itemize}
\item the boundary case in which the soft particle is adjacent to the off-shell line,

\item the case in which the soft particle is an even number one and nonadjacent to the off-shell line,

\item The case in which the soft particle is an odd number one and nonadjacent to the off-shell line.
\end{itemize}
We will prove the boundary case separately from two nonadjacent cases because of different strategies to achieve the final conclusion.

\subsubsection{The soft particle is adjacent to the off-shell line}

We choose the soft particle $i$ as $2$ or ${2n}$ which is adjacent to the off-shell line.
Since both $2$ and $2n$ are even, the $\tau^0$ term of $J(2,\dots,2n)$ will vanish eventually due to \eqref{Eq:Off-shell-Leading-Single-Soft}. In this boundary case, we actually only need to prove
\bea
 J^{(0)}(\W2 ,3,\dots, 2n)=0,\Label{Eq:Off-shell-Leading-Single-Soft1}
 \eea
  where $J^{(0)}$ stands for the $\tau^0$ order. The other case with particle $2n$ soft can be obtained via a reflection.
  The 4- and 6-point examples have already been illustrated in \secref{single-example}, and moreover, we assume that \eqref{Eq:Off-shell-Leading-Single-Soft1} is satisfied by currents $J(\W 2,\dots,2m)$ with $m<n$.

\begin{figure}
  \centering
 \includegraphics[width=1\textwidth]{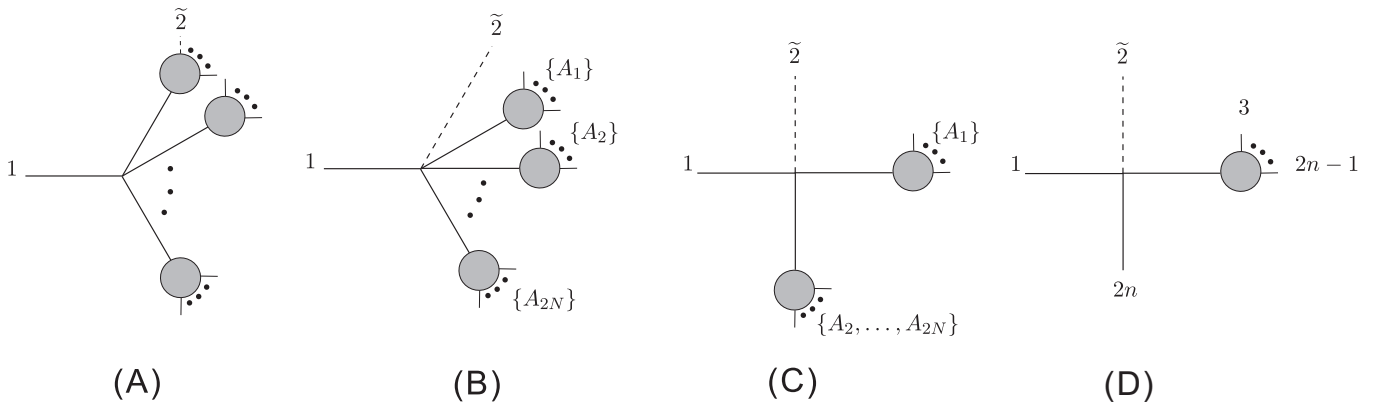}
 \caption{ Diagrams for the current $J(\W 2,\dots,2n)$ with $2$ soft. } \label{FigSingleSoft1}
\end{figure}

To prove \eqref{Eq:Off-shell-Leading-Single-Soft1}, we notice that in {the} Berends-Giele recursion of $J(\W2 ,3,\dots, 2n)$ \eqref{B-G}, the soft particle $2$ can be either in a subcurrent or connected to the off-shell line directly.
If the soft particle $2$ is in a subcurrent $J(\W 2,\dots)$ (see the diagram (A) in \figref{FigSingleSoft1}), the $\tau^0$ order of this subcurrent is zero according to our inductive assumption.
Thus the remaining diagrams are only those with $2$ connected to the off-shell line directly which can be further classified into two types.
\begin{itemize}
\item The soft particle belongs to a $2N+2$-point vertex $(1<N\leq n-1)$, as shown in diagram (B) in \figref{FigSingleSoft1}.
The leading order contribution of this type for given division $\{2,\dots, 2n\}\to\{A_1\}\dots\{A_{2N}\}$ is
\bea
\left(-{1\over 2F^2}\right)^{N}{i\over P^2_{3,2n}}i\left(\Sl_{r=1}^NP_{A_{2r}}\right)^2\prod_{s=1}^{2N}J(A_s).\Label{GenSingleSoft-B}
\eea

\item The soft particle belongs to a four-point vertex, as depicted by diagrams (C) and (D) in \figref{FigSingleSoft1}.
For a given diagram (B) in figure \ref{FigSingleSoft1}, we can always find a corresponding diagram (C) whose leading order is
\bea
&&\left(-{1\over 2F^2}\right){i\over P^2_{3,2n}(\tau=0)}i\left(\tau p+\Sl_{r=2}^{2N}P_{A_r}\right)^2\Bigg |_{\tau=0}\left(-{1\over 2F^2}\right)^{N-1}{i\over\left(\Sl_{s=2}^{2N} P_{A_{s}}\right)^2}i\left(\Sl_{t=1}^NP_{A_{2t}}\right)^2\prod\limits_{u=1}^{2N}J( A_u )\nn
&=&-\left(-{1\over 2F^2}\right)^{N}{i\over P^2_{3,2n}}i\left(\Sl_{t=1}^NP_{A_{2t}}\right)^2\prod\limits_{u=1}^{2N}J( A_u ).
\eea
This result cancels with the leading order of diagram (B), \eqref{GenSingleSoft-B}.
\end{itemize}
The only remaining contribution comes from the diagram (D).
The $\tau^0$ order of diagram (D) in \figref{FigSingleSoft1} vanishes because the leading order four-point vertex now is $(\tau p+k_{2n}^2)^2|_{\tau=0}=k_{2n}^2=0$ for massless particle $2n$.

Therefore, the $\tau^0$ order behavior of a current with a soft particle adjacent to the off-shell line is zero.

\subsubsection{The soft particle is nonadjacent to the off-shell line}
If the soft particle $i$ is nonadjacent to the off-shell line, it can also be attached to the off-shell line directly or in a subcurrent. The latter can be further classified into two classes: if the soft particle $i$ is at an even number position of a subcurrent, the $\tau^0$ order vanishes due to the inductive assumption; if the soft particle lives in an odd number position of a subcurrent, the leading order of the subcurrent is given by the non-zero term in equation \eqref{Eq:Off-shell-Leading-Single-Soft}.

\begin{figure}[H]
  \centering
 \includegraphics[width=1\textwidth]{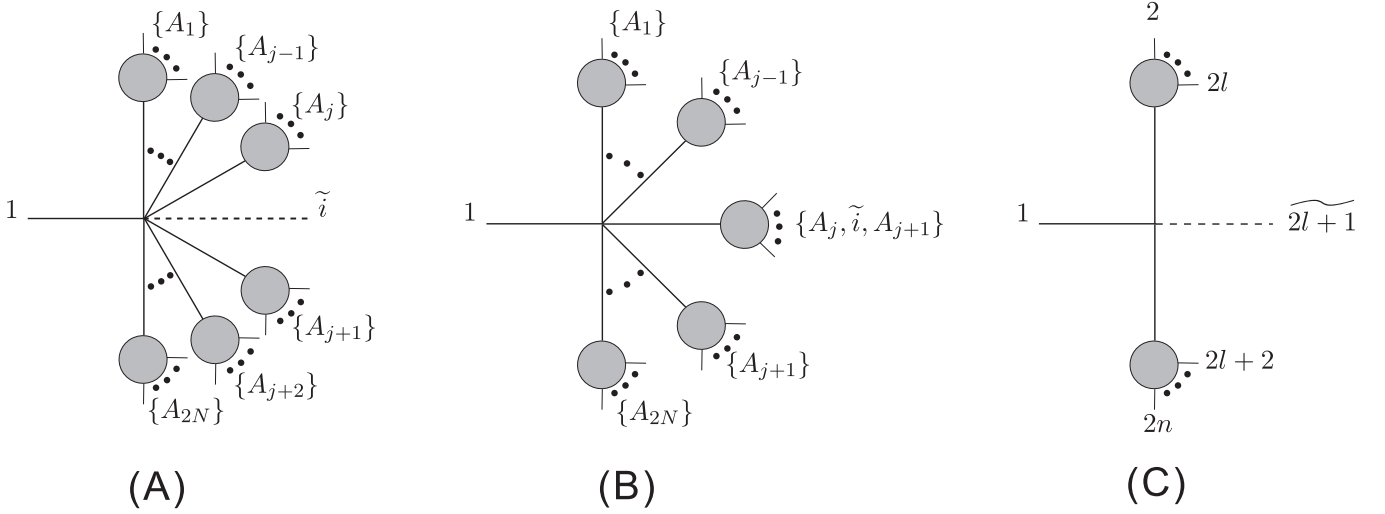}
 \caption{ Diagrams for the current $J(2,\dots,i-1,\W i,i+1,\dots,2n)$ with $i$ soft. } \label{FigSingleSoft2}
\end{figure}

As shown in \figref{FigSingleSoft2}, for a given diagram (A) with the soft particle attached to the off-shell line directly, one can always find a corresponding diagram (B) with the soft particle sitting at the odd number position of a subcurrent.
In both (A) and (B), the propagators of the off-shell lines provide a same factor
\bea
{i\over \left(\Sl_{r=1}^jP_{A_{r}}+\Sl_{s=j+1}^{2N}P_{A_{s}}\right)^2}
\eea
to the $\tau^0$ order.
The product of subcurrents in each diagram is $\prod\limits_{r=1}^jJ(A_{r})\prod\limits_{s=j+1}^{2N}J(A_{s})$, where we used our inductive assumption to express the leading order of subcurrent $J(A_{j},\W i,A_{j+1})$ in diagram (B) as $(1/2F^2) J(A_{j})J(A_{j+1})$.

\begin{itemize}
\item If $i=2l$, which means we are considering $J(2,\dots,\W{2l}, \dots, 2n)$, the $j$ in the diagrams (A) and (B) must be an even number. The leading order of the vertex in diagram (A) in \figref{FigSingleSoft2} reads
\bea
&&\left(-{1\over 2F^2}\right)^Ni\left(\Sl_{u=0}^{{j-2\over 2}}P_{A_{2u+1}}+\tau p+\Sl_{v={j+2\over 2}}^NP_{A_{2v}}\right)^2\Bigg|_{\tau=0}\nn
&=&\left(-{1\over 2F^2}\right)^Ni\left(\Sl_{u=0}^{{j-2\over 2}}P_{A_{2u+1}}+\Sl_{v={j+2\over 2}}^NP_{A_{2v}}\right)^2,\Label{single-even-A}
\eea
while the leading order of the vertex in diagram (B) in \figref{FigSingleSoft2} reads
\bea
&&\left(1\over 2F^2\right)\left(-{1\over 2F^2}\right)^{N-1}i\left(\Sl_{u=0}^{{j-2\over 2}}P_{A_{2u+1}}+\Sl_{v={j+2\over 2}}^NP_{A_{2v}}\right)^2.
\Label{single-even-B}
\eea
The isolated $\left(1\over 2F^2\right)$ in \eqref{single-even-B} comes from the $\tau^0$ order expression of $J(\{A_{j}\},\W i,\{A_{j+1}\})$ which is  mentioned before.
The results from \eqref{single-even-A} and \eqref{single-even-B} cancel with each other exactly once given the same division in diagrams (A) and (B).

\item If $i=2l+1$, which means we are considering $J(2,\dots,\W{2l+1}, \dots, 2n)$, $j$ has to be an odd number. The leading order of the vertex in (A) reads
\bea
&&\left(-{1\over 2F^2}\right)^Ni\left(\Sl_{u=0}^{{j-1\over 2}}P_{A_{2u+1}}+\Sl_{v={j+1\over 2}}^NP_{A_{2v}}\right)^2,
\eea
while the leading order of the vertex in (B) reads
\bea
&&\left({1\over 2F^2}\right)\left(-{1\over 2F^2}\right)^{N-1}i\left(\Sl_{u=0}^{{j-1\over 2}}P_{A_{2u+1}}+\tau p+\Sl_{v={j+1\over 2}}^NP_{A_{2v}}\right)^2\Bigg|_{\tau=0}\nn
&=&-\left(-{1\over 2F^2}\right)^{N}i\left(\Sl_{u=0}^{{j-1\over 2}}P_{A_{2u+1}}+\Sl_{v={j+1\over 2}}^NP_{A_{2v}}\right)^2.
\eea
Again, the leading contributions from the vertices of diagram (A) and (B) in figure \ref{FigSingleSoft2} cancel with each other.
\end{itemize}

The diagram (C) in \figref{FigSingleSoft2} as a special case of (A) contains a four-point vertex and needs to be considered independently. In this case, $i=2l+1$ is required and no corresponding diagram (B) can be found to cancel the leading term of (C).
The $\tau^0$ order of the diagram (C) in figure \ref{FigSingleSoft2} gives
\bea
&&\left(-{1\over 2F^2}\right){i\over P_{2,2n}^2({k_{2l+1}=0})}i\left(\Sl_{r=2}^{2l}k_r+\Sl_{s=2l+2}^{2n}k_s\right)^2J(2,\dots,2l)J(2l+1,\dots,2n)\nn
&=&\left(1\over 2F^2\right)J(2,\dots,2l)J(2l+1,\dots,2n).
\eea
Thus the soft behavior \eqref{Eq:Off-shell-Leading-Single-Soft} has been proven.

\section{Double soft behavior of currents with two adjacent soft particles}\label{sec:offshell-double-soft}
In this section, we will discuss the leading and subleading order behaviors of a current $J(\dots,\W i,\W {i+1},\dots)$ where the momenta of two adjacent particles $i$, $i+1$ tend to zero. In general, $i$ in $J(\dots,\W i,\W {i+1},\dots)$ can be either even or odd.  A current with an even $i$ can always be reflected to a current with an odd $i$ via \eqref{eq:reflection}. Thus we only need to prove one case. In this section, we prove the leading and  subleading double soft behaviors of $J(\dots,\W i,\W {i+1},\dots)$ with even $i$. Before the general proof, we first illustrate explicit calculations on the double soft behaviors of four- and six- point currents.

\subsection{Four-point current with two soft particles}
We now consider the four-point current $J(\W 2,\W 3, 4)$ where the soft particle $2$ is adjacent to the off-shell line. The momenta of the soft particles $2$ and $3$ are $k_2=\tau p$ and $k_3=\tau q$ respectively. The current can be explicitly written as
\bea
J(\W 2,\W 3,4)=\left(-{1\over 2F^2}\right){i\over (\tau p+\tau q+k_4)^2}2i\tau k_4\cdot p&=&\left({1\over 2F^2}\right){k_4\cdot p\over k_4\cdot (p+q)+\tau p\cdot q}.
\eea
Here we have used on-shell conditions for external particles $k_2^2=p^2=q^2=0$. After expanding the current with respect to $\tau$, we find the leading and subleading terms under $\tau\to 0$ is
\bea
J(\W 2,\W 3, 4)=\tau^0\left({1\over 2F^2}\right){k_4\cdot p \over k_4\cdot (p+q)}+\tau^1\left(-{1\over 2F^2}\right)(p\cdot q){k_4\cdot p\over( k_4\cdot (p+q))^2}+\mathcal{O}(\tau^2).
\eea
Recalling that $J(4)=1$, ${\cal J}^{\mu\nu}_4[J(4)]$ has to vanish. Thus the above equation gives the expected leading and subleading terms of double soft behavior of the four-point current, as proposed in \eqref{Off-shell-Leading-Subleading-Double-Soft0}.

\subsection{Six-point current with two soft particles}
The six-point current with two soft particles $i$, ${i+1}$ is more complicated but can present most properties in an arbitrary-point current. As already discussed at the beginning of this section, we only need to consider even $i$. For the six-point case, $i$ can be chosen as $2$ or $4$, which corresponds to the following two classes according to the position of the soft particle $i$
\begin{itemize}
\item[] (a) $i=2$, the soft particle $i$ is adjacent to the off-shell line, i.e., $J(\W 2,\W 3, 4, 5, 6)$,
\item[] (b) $i=4$, neither $i$ nor ${i+1}$ is adjacent to the off-shell line, i.e., $J(2, 3, \W 4, \W 5, 6)$.
\end{itemize}
These two classes for six-point current are very simple but largely show the logics in a general case.

\subsubsection*{Take the particles $2$ and $3$ as soft}
%

The six-point current $J(\W 2,\W 3,4, 5, 6)$ with $2$ and $3$ soft is
\bea
J(\W 2,\W 3,4, 5, 6)={i\over P^2_{2,6}(k_2=\tau p,k_3=\tau q)}i\left[T_A(\W 2,\W 3)+T_B(\W 2,\W 3)+T_C(\W 2,\W 3)+T_D(\W 2,\W 3)\right],\Label{EQ-Off-shell-double-soft-6pt-Adj}
\eea
where $T_A(\W 2,\W 3)$, $T_B(\W 2,\W 3)$, $T_C(\W 2,\W 3)$ and $T_D(\W 2,\W 3)$ come from (excluding the propagator of the off-shell line)  the diagrams (A), (B), (C) and (D) in \figref{FigAll6PtDiagrams}. While $\tau\to 0$, the propagator of the off-shell line is expanded as
\bea
{1\over P^2_{2,6}(k_2=\tau p,k_3=\tau q)}=\tau^0{1\over P_{4,6}^2}-\tau^1{2P_{4,6}\cdot (p+q)\over (P_{4,6}^2)^2}+\mathcal{O}(\tau^2).~~\Label{Propagator6pt1}
\eea
Inserting the leading and subleading order expressions of $J(\W 2,\W 3,4)$ as well as the leading order behavior of $J(\W 3,4,5)$ into the diagrams (B) and (C) in \figref{FigAll6PtDiagrams} respectively, we calculate complete contributions (apart from the propagator of the off-shell line) from \figref{FigAll6PtDiagrams} (see \tabref{Table1}).

\begin{table}[H]{\centering
\begin{tabular}{c|c|c}
  \hline
    & $\tau^0$ & $\tau^1$ \\ \hline \hline
  $T_A(\W 2,\W 3)$ & $P_{4,6}^2J(4,5,6)$  & $2(k_4+k_6)\cdot p $ \\ \hline
   $ T_B(\W 2,\W 3)$ & ${k_4\cdot p\over k_4\cdot(p+q)}\left(-P_{4,6}^2\right)J(4,5,6)$  & $-(p\cdot q){k_4\cdot p\over (k_4\cdot(p+q))^2}\left(-P_{4,6}^2\right)J(4,5,6)-2(k_4+k_6)\cdot(p+q){k_4\cdot p\over k_4\cdot(p+q)}$  \\ \hline
    $T_C(\W 2,\W 3)$ & $0$  & $0$  \\ \hline
    $T_D(\W 2,\W 3)$ & $-P^2_{4,6}J(4,5,6)$  & $-2P_{4,6}\cdot p J(4,5,6)$  \\
  \hline
\end{tabular}\caption{The $\tau^0$ and $\tau^1$ order terms of diagrams (A), (B), (C) and (D) in \figref{FigAll6PtDiagrams} with particles $2$ and $3$ soft. The constant ${1\over 2F^2}$ has been set to 1 for convenience. The expression of the subcurrent $J(4,5,6)=(-1){i\over P^2_{4,6}}i(k_4+k_6)^2$ has been taken into account. On-shell conditions are considered in $T_C(\W 2,\W 3)$.}\Label{Table1}}
\end{table}

\begin{itemize}
\item \emph{\bf{Leading order}}

From \tabref{Table1}, we observe that the $\tau^0$  terms of $T_A(\W 2,\W 3)$ and $T_D(\W 2,\W 3)$ cancel with each other. The $\tau^0$ term of $T_C(\W 2,\W 3)$ vanishes because it contains a subcurrent $J(\W 3,4,5)$ whose leading order has been proven to be zero in the previous section.
Only the diagram (B) in \figref{FigAll6PtDiagrams} contains a nonzero contribution. Substituting the $\tau^0$  terms of the propagator \eqref{Propagator6pt1} and  $T_B(\W 2,\W 3)$ into  \eqref{EQ-Off-shell-double-soft-6pt-Adj}, we obtain the expected leading order double soft behavior
\bea
J^{(0)}(\W 2,\W 3,4,5,6)=\left(1\over 2F^2\right){k_4\cdot p\over k_4\cdot(p+q)}J(4,5,6)=S^{(0)}_{2,3}J(4,5,6),\Label{Eq-Six-Point-(23)-Leading}
\eea
where we have put the factor $\left(1\over 2F^2\right)$ back.

\item \emph{\bf{Subleading order}}

Through \tabref{Table1} and \eqref{Propagator6pt1}, we read off the coefficient of $\tau^1$ in \eqref{EQ-Off-shell-double-soft-6pt-Adj}
\bea
J^{(1)}(\W 2,\W 3,4,5,6)&=&-{2P_{4,6}\cdot (p+q)\over P_{4,6}^2}\left(1\over 2F^2\right){k_4\cdot p\over k_4\cdot(p+q)}J(4,5,6)\nn
&&+{1\over P_{4,6}^2}\left({1\over 2F^2}\right)^2\Biggl[-2(k_4+k_6)\cdot p+2(k_4+k_6)\cdot(p+q){k_4\cdot p\over k_4\cdot(p+q)}\Biggr]\nn
&&+{1\over P_{4,6}^2}\left({1\over 2F^2}\right)\Biggl[-(p\cdot q){k_4\cdot p\over (k_4\cdot(p+q))^2}P_{4,6}^2J(4,5,6)+2P_{4,6}\cdot p\, J(4,5,6)\Biggr]. \nonumber\\
\Label{Eq-Six-Point-(23)-subLeading}
\eea
Using the following property
\bea
{k_r\cdot p\over k_r\cdot(p+q)}2K\cdot (p+q)-2K\cdot p&=&{p_{\mu}q_{\nu}\over k_r\cdot(p+q)}{\cal J}_r^{\mu\nu} K^2, ~~\Label{property1}
\eea
we simplify the second line of \eqref{Eq-Six-Point-(23)-subLeading} as well as the sum of the first line and the second term of the third line there.
Then the subleading order of  \eqref{Eq-Six-Point-(23)-subLeading} becomes
\bea
J^{(1)}(\W 2,\W 3,4,5,6)=\left(-{1\over 2F^2}\right)\left[(p\cdot q){k_4\cdot p\over (k_4\cdot(p+q))^2}+{q_{\mu}p_{\nu}{\cal J}_4^{\mu\nu}\over k_4\cdot(p+q)}\right]J(4,5,6)=S^{(1)}_{2,3}J(4,5,6).\nonumber\\
\eea
Thus the expected subleading order double soft behavior is also achieved.
\end{itemize}
%

\subsubsection*{Take the particles $4$ and $5$ as soft}

If the particles $4$ and $5$ are soft,  we have
\bea
J(2,3,\W 4,\W 5,6)={i\over P^2_{2,6}(k_4=\tau p,k_5=\tau q)}i\left[T_A(\W 4,\W 5)+T_B(\W 4,\W 5)+T_C(\W 4,\W 5)+T_D(\W 4,\W 5)\right],\Label{EQ-Off-shell-double-soft-6pt-nonAdj}
\eea
where the propagator can be expanded as
\bea
{1\over P^2_{2,6}(k_4=\tau p,k_5=\tau q)}=\tau^0{1\over (k_2+k_3+k_6)^2}-\tau^1 {2(k_2+k_3+k_6)\cdot (p+q)\over ((k_2+k_3+k_6)^2)^2}+\mathcal{O}(\tau^2).~~\Label{Propagator6pt2}
\eea
Substituting the soft behaviors of  $J(2,3,\W 4)$, $J(3,\W 4,\W 5)$ and $J(\W 4,\W 5, 6)$ into contributions from the diagrams (B), (C) and (D) in \figref{FigAll6PtDiagrams} respectively and expanding the vertices into series of $\tau$, we find the leading and subleading contributions (apart from the propagator of the off-shell line) from all diagrams (A), (B), (C) and (D) (see \tabref{Table2}).

\begin{table}[H]{\centering
\begin{tabular}{c|c|c}
  \hline
    & $\tau^0$ & $\tau^1$ \\ \hline \hline
  $T_A(\W 4,\W 5)$ & $(k_2+k_3+k_6)^2J(2,3,6)$  & $2(k_2+k_6)\cdot p $ \\ \hline
   $ T_B(\W 4,\W 5)$ & $0$  & $-(k_2+k_3+k_6)^2J^{(1)}(2,3,\W 4)$  \\ \hline
    $T_C(\W 4,\W 5)$ & ${k_3\cdot q \over k_3\cdot (p+q)}\left[-(k_2+k_3+k_6)^2\right]J(2,3,6)$  & $-(p\cdot q){k_3\cdot q\over( k_3\cdot (p+q))^2}\left[-(k_2+k_3+k_6)^2\right]J(2,3,6)$  \\ \hline
    $T_D(\W 4,\W 5)$ & ${k_6\cdot p \over k_6\cdot (p+q)}\left[-(k_2+k_3+k_6)^2\right]J(2,3,6)$  & $
                                                                                                     \begin{array}{cc}
                                                                                                      &-2(k_2+k_6)\cdot (p+q){k_6\cdot p \over k_6\cdot (p+q)}~~~~~~~~~~~~~~~~~\\
                                                                                                      &+(p\cdot q){k_6\cdot p\over( k_6\cdot (p+q))^2}\left[-(k_2+k_3+k_6)^2\right]J(2,3,6) \\
                                                                                                     \end{array}$
     \\
  \hline
\end{tabular}\caption{The $\tau^0$ and $\tau^1$ order terms of the diagrams (A), (B), (C), (D) in figure \ref{FigAll6PtDiagrams} with particles $4$, $5$ soft. As in \tabref{Table1}, the constant ${1\over 2F^2}$ has been set to 1. The expression of the subcurrent $J(2,3,6)=(-1){i\over (k_2+k_3+k_6)^2}i(k_2+k_6)^2$ has been taken into account. On-shell conditions are used to derive $T_B(\W 4,\W 5)$. }\Label{Table2}}
\end{table}

\begin{itemize}
\item \emph{\bf{Leading order}}

Plugging all $\tau^0$ terms shown in table \ref{Table2} and the $\tau^0$ term of equation \eqref{Propagator6pt2} into \eqref{EQ-Off-shell-double-soft-6pt-nonAdj}, we immediately obtain
\bea
J^{(0)}(2,3,\W 4,\W 5,6)={1\over 2}\left(-{1\over 2F^2}\right)\left[{k_6\cdot(q-p)\over k_6\cdot(q+p)}+{k_3\cdot(p-q)\over k_3\cdot(p+q)}\right]J(2,3,4)=S^{(0)}_{4,5}J(2,3,4),
\eea
where we have put $\left({1\over 2F^2}\right)$ back. This is nothing but the expected leading order double soft behavior \eqref{Off-shell-Leading-Subleading-Double-Soft1}.

\item \emph{\bf {Subleading order }}

The subleading order of $J(2,3,\W 4,\W 5,6)$  can be arranged as
\bea
&&J^{(1)}(2,3,\W 4,\W 5,6) ~~~~\Label{Eq-Six-Point-(45)-subLeading}\\
&=&-{2(k_2+k_3+k_6)\cdot (p+q)\over (k_2+k_3+k_6)^2}\left({1\over 2F^2}\right)\left[{k_3\cdot q \over k_3\cdot (p+q)}+{k_6\cdot p \over k_6\cdot (p+q)}-1\right]J(2,3,6)\nn
&&+{1\over (k_2+k_3+k_6)^2}\left({1\over 2F^2}\right)^2\left[-2(k_2+k_6)\cdot p+2(k_2+k_6)\cdot (p+q){k_6\cdot p \over k_6\cdot (p+q)}\right]\nn
&&+\left({1\over 2F^2}\right)\Biggl[-(p\cdot q)\left({k_3\cdot q\over( k_3\cdot (p+q))^2}
+{k_6\cdot p\over( k_6\cdot (p+q))^2}\right)J(2,3,6)+J^{(1)}(2,3,\W 4)\Biggr].\nonumber
\eea
Here, the first line on the right hand side is a product of the $\tau^1$ term of the propagator in equation \eqref{Propagator6pt2} and the sum of $\tau^0$ terms of the $T$'s in \tabref{Table2}, while the sum of the second and the third lines corresponds to the product of the $\tau^1$ term of the propagator \eqref{Propagator6pt2}  and the sum of $\tau^0$ terms of the $T$'s in \tabref{Table2}. Applying the property \eqref{property1} to the first line and noticing that $2K\cdot (p+q)=2K\cdot p+2K\cdot q$, we have
\bea
-\left({1\over 2F^2}\right){1\over (k_2+k_3+k_6)^2}\left[\left({p_{\mu}q_{\nu}{\cal J}_6^{\mu\nu}\over k_6\cdot(p+q) }+{q_{\mu}p_{\nu}{\cal J}_3^{\mu\nu}\over k_3\cdot(p+q) }\right)(k_2+k_3+k_6)^2\right]J(2,3,6).
\eea
The second line of \eqref{Eq-Six-Point-(45)-subLeading} can also be simplified as
\bea
&&\left({1\over 2F^2}\right){1\over (k_2+k_3+k_6)^2}\left[{p_{\mu}q_{\nu}{\cal J}_6^{\mu\nu}\over k_6\cdot(p+q) }\left((k_2+k_3+k_6)^2J(2,3,6)\right)\right]\\
&=&\left({1\over 2F^2}\right){1\over (k_2+k_3+k_6)^2}\left[\left({p_{\mu}q_{\nu}{\cal J}_6^{\mu\nu}\over k_6\cdot(p+q) }+{q_{\mu}p_{\nu}{\cal J}_3^{\mu\nu}\over k_3\cdot(p+q) }\right)\left((k_2+k_3+k_6)^2J(2,3,6)\right)\right],\nonumber
\eea
where we used the explicit form of $J(2,3,6)$ as well as the property \eqref{property1} again, and also added ${\cal J}_3^{\mu\nu}\left((k_2+k_3+k_6)^2J(2,3,6)\right)$ to the final results due to $(k_2+k_3+k_6)^2J(2,3,6)=({1\over 2F^2})(k_2+k_6)^2$ and ${\cal J}_3^{\mu\nu}\left((k_2+k_3+k_6)^2J(2,3,6)\right)=0$.

Finally, we achieve the subleading order behavior of $J^{(1)}(2,3,\W 4,\W 5,6)$ as
\bea
J^{(1)}(2,3,\W 4,\W 5,6)&=&\left(-{1\over 2F^2}\right)\Biggl[(p\cdot q)\left({k_3\cdot q\over( k_3\cdot (p+q))^2}
+{k_6\cdot p\over( k_6\cdot (p+q))^2}\right)\\
&&~~~~~~~~~~+\left({q_{\mu}p_{\nu}{\cal J}_6^{\mu\nu}\over k_6\cdot(q+p) }+{p_{\mu}q_{\nu}{\cal J}_3^{\mu\nu}\over k_3\cdot(p+q) }\right)\Biggr]J(2,3,6)+\left({1\over 2F^2}\right)J^{(1)}(2,3,\W 4)J(6)\nn
&=&S^{(1)}_{4,5}J(2,3,6)+\left({1\over 2F^2}\right)J^{(1)}(2,3,\W 4)J(6),
\eea
which is the expected subleading behavior of $J^{(1)}(2,3,\W 4,\W 5,6)$ with $4$ and $5$ as soft.

\end{itemize}
%


\subsection{General proof}
Enlightened by the explicit four-point and six-point examples, we turn to prove the double soft behavior of an arbitrary-point current $J(2, \dots,\W{2l},\W{2l+1},\dots, 2n)$ now.
As done in the six-point example, we first consider the boundary case where one of the soft particles is adjacent to the off-shell line, i.e., $l=1$, and then  study the case with no soft particle adjacent to the off-shell line, i.e., $l>1$.

\subsubsection{One soft particle is adjacent to the off-shell line}
\begin{figure}
  \centering
 \includegraphics[width=0.7\textwidth]{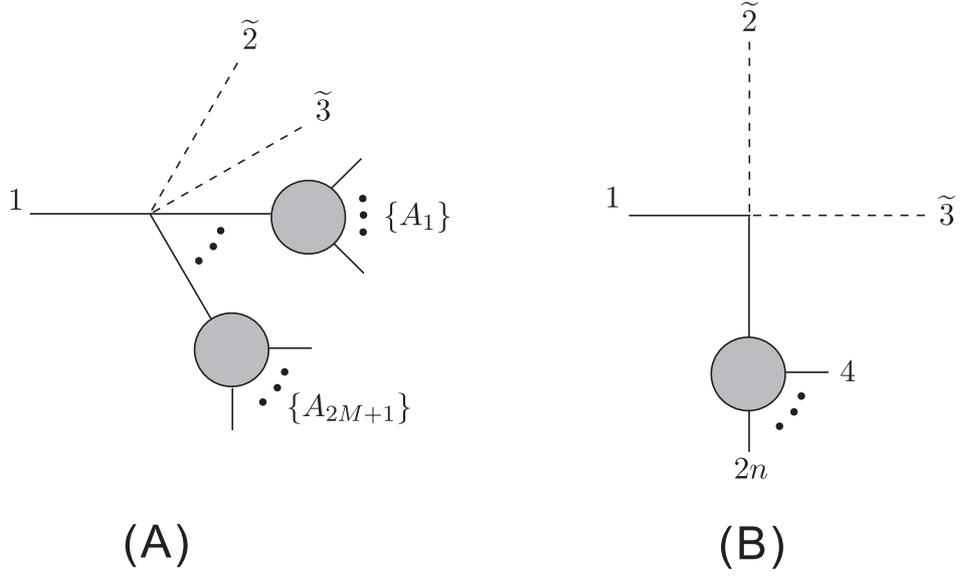}
 \caption{A typical diagram with the soft particles $2$ and $3$ attached to the off-shell line is given by (A). (B) is the boundary case with the off-shell line connected to a four-point vertex.} \label{FigGenAdjType1}
\end{figure}

\begin{figure}
  \centering
 \includegraphics[width=1\textwidth]{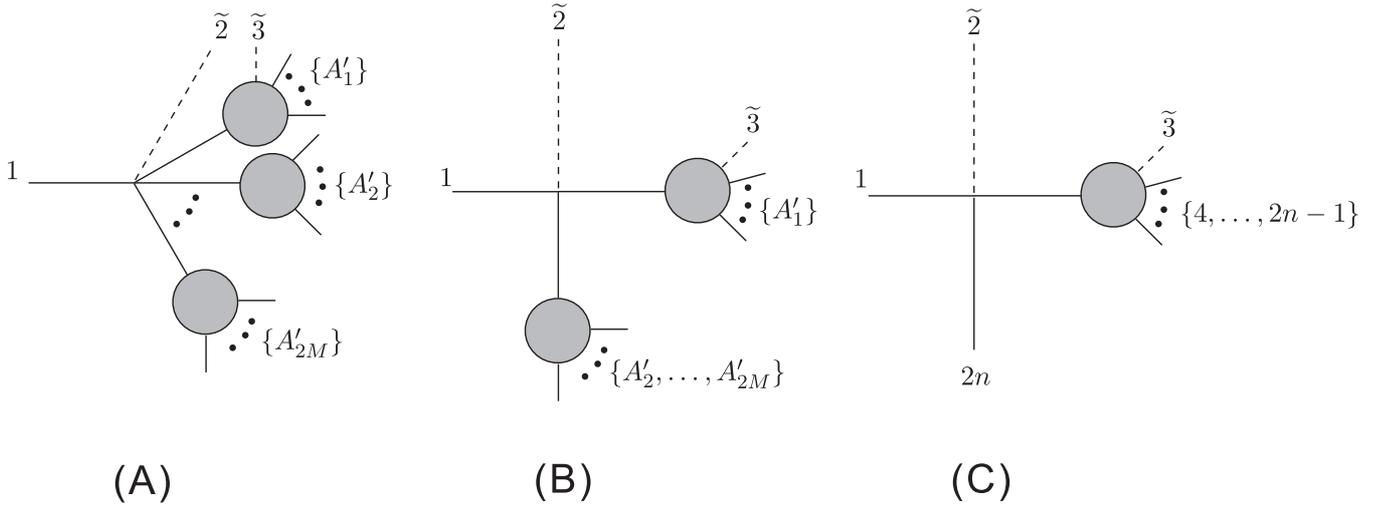}
 \caption{A typical diagram with the soft particles $2$ attached to the off-shell line and the soft particle $3$ in a subcurrent is given by (A). (B) and (C) are two boundary cases with the off-shell line connected to a four-point vertex.} \label{FigGenAdjType2}
\end{figure}

\begin{figure}
  \centering
 \includegraphics[width=0.3\textwidth]{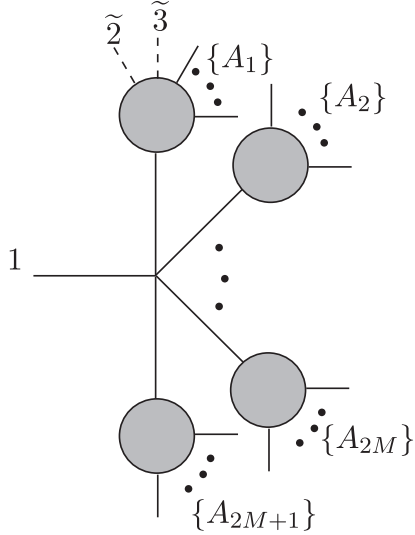}
 \caption{A typical diagram with both soft particles $2$ and $3$ in the same subcurrent.} \label{FigGenAdjType3}
\end{figure}

Let us start from the double soft behavior of the current $J(\W 2,\W 3,\dots,2n)$ where the soft particle $2$ is adjacent to the off-shell line.
We express $J(\W 2,\W 3,\dots,2n)$ in terms of products of lower-point subcurrents by Berends-Giele recursion, then classify all diagrams emerging from Berends-Giele recursion into three types according to the positions of the soft particles:
\begin{itemize}
\item[]\textit{{type-1}} diagrams with both soft particles $2$ and $3$ attached to the off-shell line, as shown in \figref{FigGenAdjType1},
\item[] \textit{{type-2}} diagrams with the soft particle $2$ attached to the off-shell line and the soft particle $3$ in a subcurrent, as shown in \figref{FigGenAdjType2},
\item[]\textit{{type-3}} diagrams with both soft particles $2$ and $3$ in the same subcurrent, as shown in \figref{FigGenAdjType3}.
\end{itemize}
With this classification, the current $J(\W 2,\W 3, \dots, 2n)$ is given by
\bea
J(\W 2,\W 3, \dots, 2n)&=&{i\over P^2_{2,2n}(k_2=\tau p,k_3=\tau q)} i\left[T^{\text{A}}_{\text{Type-1}}(\tau)+T^{\text{A}}_{\text{Type-2}}(\tau)+T^{\text{A}}_{\text{Type-3}}(\tau)\right],\Label{EqGenAdj}
\eea
where $T^{\text{A}}_{\text{Type-1}}(\tau)$, $T^{\text{A}}_{\text{Type-2}}(\tau)$ and $T^{\text{A}}_{\text{Type-3}}(\tau)$ \footnote{The superscript `A' implies that the soft particle $2$ is adjacent to the off-shell line; readers will see another superscript `N' in section \ref{subsubsect-nonadj} and \appref{app:appendix-nonadj}.} denote the contributions from different types (apart from the propagator of the off-shell line) .
When $\tau\to 0$, the propagator can be easily expanded as
\bea
{1\over P^2_{2,2n}(k_2=\tau p,k_3=\tau q)}=\tau^0{1\over P^2_{4,2n}}-\tau^1{2P_{4,2n}\cdot (p+q)\over \left(P^2_{4,2n}\right)^2}+\mathcal{O}(\tau^2).\Label{PropagatorGenAdj}
\eea
The contributions from these three types will be considered one by one.


\begin{itemize}
\item \textit{{Type-1}} Diagrams of this type are typically shown in \figref{FigGenAdjType1}. The sum of  all diagrams with  $M>0$ (see diagram (A) in \figref{FigGenAdjType1}) is
expanded as
\bea
T^{\text{A}}_{\text{Type-1(A)}}(\tau)&=&\tau^0\Sl_{\{4,\dots,2n\}\to\{A_1,\dots,A_{2M+1}\}}\left(-{1\over 2F^2}\right)^{M+1}\left(\Sl_{r=0}^MP_{A_{2r+1}}\right)^2\prod\limits_{j=1}^{2M+1}J\left(A_{j}\right)\Label{EqGenAdjType1A}\nn
&+&\tau^1\Sl_{\{4,\dots,2n\}\to\{A_1,\dots,A_{2M+1}\}}\left(-{1\over 2F^2}\right)^{M+1}2\left(\Sl_{r=0}^MP_{A_{2r+1}}\right)\cdot p\prod\limits_{j=1}^{2M+1}J\left(A_{j}\right).
\eea
The special case with $M=0$ is depicted by the diagram (B) in \figref{FigGenAdjType1} and its first two orders are
\bea
T^{\text{A}}_{\text{Type-1(B)}}(\tau)=\tau^0\left(-{1\over 2F^2}\right)P^2_{4,2n}J(4,\dots,2n)+\tau^1\left(-{1\over 2F^2}\right)2P_{4,2n}\cdot p\, J(4,\dots,2n).\Label{EqGenAdjType1B}
\eea
Considering the Berends-Giele recursion of the current $J(4,\dots,2n)$, we find that the $\tau^0$ terms of \eqref{EqGenAdjType1A} and \eqref{EqGenAdjType1B} cancel with each other. Thus only the $\tau^1$ term survives from summing over all type-1 diagrams and is expressed as
\bea
T^{\text{A}}_{\text{Type-1}}(\tau)&=&\tau^1\Biggl[\Sl_{\{4,\dots,2n\}\to\{A_1,\dots,A_{2M+1}\}}\left(-{1\over 2F^2}\right)^{M+1}2\left(\Sl_{r=0}^MP_{A_{2r+1}}\right)\cdot p\prod\limits_{j=1}^{2M+1}J\left(A_{j}\right)\nn
&&~~~~+\left(-{1\over 2F^2}\right)2P_{4,2n}\cdot p J(4,\dots,2n)\Biggr].\Label{EqGenAdjType1}
\eea

\item \textit{{Type-2}} Diagrams with the soft particle $2$ attached to the off-shell line and the soft particle $3$ in a subcurrent can be further classified into (A), (B) and (C) as shown in figure  \ref{FigGenAdjType2}.
\begin{itemize}
  \item Diagram (A) represents the case with the off-shell line attached to a $2M+2$-point $(M>1)$ vertex.
  \item Diagram (B) stands for the case in which the off-shell line is attached to a $4$-point vertex with a combination of $\{A^\prime_2, \dots, A^\prime_{2M}\}$ as the last subcurrent.
  \item Diagram (C) is the case that the off-shell line is attached to a $4$-point vertex whose other three legs are connected to the soft particle $2$, the subcurrent $J(\W 3,4,\dots, 2n-1)$ and the particle $2n$.
\end{itemize}
Apparently, (C) is the boundary case of (B) while setting $\{A_2^\prime,\dots,A_{2M}^\prime\}=\{2n\}$, while both (C) and (B) are the boundary cases of (A) with $M=1$.

We now show that all the leading and subleading terms in Type-2 diagrams cancel out.
Given the division $\{4,\dots, 2n\}\to\{A_1^\prime\},\dots,\{A_{2M}^\prime\}$ in which $\{A_1^\prime\}$ can only contain particles of an even number while $\{A_{j}\}(j>1)$ can only contain particles of an odd number, the contribution from the diagram (A) in \figref{FigGenAdjType2} can be written explicitly as
    \bea
  &&\left(-{1\over 2F^2}\right)^M\left(\tau p+\Sl_{r=1}^MP_{A_{2r}^\prime}\right)^2 J(\W 3,A_1^\prime)\prod\limits_{j=2}^{2M}J(A_{j}^\prime)\Label{EqGenAdjType2A}\nn
  &=&\tau^1\left(-{1\over 2F^2}\right)^M\left(\Sl_{r=1}^MP_{A_{2r}^\prime}\right)^2 J^{(1)}(\W 3,A_1^\prime)\prod\limits_{j=2}^{2M}J(A_{j}^\prime)+\mathcal{O}(\tau^2),
    \eea
where we used the fact that $J^{(0)}(\W 3,\dots)$ with a single soft particle $3$ (the even $i$ case of the single soft behavior \eqref{Eq:Off-shell-Leading-Single-Soft}) has to vanish.
Correspondingly, the diagram (B) contributes
\bea
 \left(-{1\over 2F^2}\right)\left(\tau p+\Sl_{r=2}^{2M}P_{A_{r}^\prime}\right)^2J(\W 3,A_1^\prime)\left[\left(-{1\over 2F^2}\right)^{M-1}{i\over \left(\Sl_{r=2}^{2M}P_{A_{r}^\prime}\right)^2}i\left(\Sl_{t=1}^{M}P_{A_{2t}^\prime}\right)^2\prod\limits_{r=2}^{2M}J(A_{r}^\prime)\right]\Label{EqGenAdjType2B}
\eea
to the division $\{A_1^\prime\},\dots,\{A_{2M}^\prime\}$.
Those factors in the brackets of \eqref{EqGenAdjType2B} come from the Berends-Giele recursion of the subcurrent $J(A_2^\prime,\dots,A_{2M}^\prime)$ in the diagram (B) of \figref{FigGenAdjType2}.
Expanding \eqref{EqGenAdjType2B} with respect to $\tau$ and using $J^{(0)}(\W 3,\dots )=0$, we find that the $\tau^0$ order of \eqref{EqGenAdjType2B} vanishes.
The subleading order of \eqref{EqGenAdjType2B} is identical to the negative of \eqref{EqGenAdjType2A}.
Thus the subleading terms from diagrams (A) and (B) cancel out.
The only contribution isolated from (A) and (B) is the diagram (C), because no counter diagram in (A) can be used to cancel with (C).
The diagram (C)  is written as %
\bea
&&\left(-{1\over 2F^2}\right)(\tau p+k_{2n})^2J(\W 3, 4, \dots, 2n-1).\Label{EqGenAdjType2C}
\eea
The $\tau^0$ and $\tau^1$ terms will vanish due to the on-shell condition $k_{2n}^2=0$ and $J^{(0)}(\W 3,A_1^\prime,\dots,A_{2n-1}^\prime)=0$.

\item \textit{{Type-3}}
Figure \ref{FigGenAdjType3} presents this type of contribution.
Given division $\{4,\dots,2n\}\to\{A_1\},\dots,\{A_{2M+1}\}$, we can expand the vertex $(\tau p+\tau q+\Sl_{r=0}^M P_{A_{2r+1}})^2$ with respect to $\tau$ and also express the subcurrent $J(\W 2,\W 3, A_1)$ due to the inductive assumption of the double soft behavior for lower-point current.
After summing over all possible divisions, we obtain
    \bea
    %
    T^{\text{A}}_{\text{Type-3}}(\tau)&=&\tau^0S^{(0)}_{2,3}\left(-P^2_{4,2n}\right)J(4,\dots,2n)\nn
    &+&\tau^1\Sl_{\{4,\dots,2n\}\to\{A_1,\dots,A_{2M+1}\}}\left(-{1\over 2F^2}\right)^M\Biggl[2\left(\Sl_{r=0}^M P_{A_{2r+1}}\right)\cdot(p+q)\,S^{(0)}_{2,3}\,J(A_1)\prod\limits_{s=2}^{2M+1}J(A_s)\nn
    &&~~~~~~~~~~~~~~~~~~~~~~~~~~~+\left(\Sl_{r=0}^MP_{A_{2r+1}}\right)^2\left(S^{(1)}_{2,3}J(A_1)\right)\prod\limits_{s=2}^{2M+1}J(A_s)\Biggr],\Label{EqGenAdjType3}
    \eea
where  the Berends-Giele recursion of $J(4,\dots,2n)$ has been used.
\end{itemize}

We already observed the cancellation among all diagrams of type-2, and now we sum  $T^{\text{A}}_{\text{Type-1}}(\tau)$ in \eqref{EqGenAdjType1} and $T^{\text{A}}_{\text{Type-3}}(\tau)$ in \eqref{EqGenAdjType3} together. Taking the propagator \eqref{PropagatorGenAdj} into account,
we obtain the leading and subleading terms of \eqref{EqGenAdj} as follows.
\begin{itemize}
\item \textit{Leading order}\\
The nontrivial $\tau^0$ contribution only comes from \eqref{EqGenAdjType3} is written as
\bea
{i\over P^2_{4,2n}}iS^{(0)}_{2,3}\left(-P^2_{4,2n}\right)J(4,\dots,2n)=S^{(0)}_{2,3}J(4,\dots,2n). \Label{EqGenAdjLeading}
\eea
\item \textit{Subleading order}\\
The $\tau^1$ coefficient can be rearranged as
\bea
&&S^{(0)}_{2,3}{2P_{4,2n}\cdot (p+q)\over P^2_{4,2n}}\left(-1\right)J(4,\dots,2n)+\left({1\over 2F^2}\right){{2P_{4,2n}\cdot p\over P^2_{4,2n}}} J(4,\dots,2n)\Label{EqGenAdjSubLeading}\nn
&+&\Sl_{\substack{\{4,\dots,2n\}\\\to\{A_1,\dots,A_{2M+1}\}}}\left({-{1\over P^2_{4,2n}}}\right)\left(-{1\over 2F^2}\right)^{M}\,\nn
&&\qquad\times\Biggl[-\left({1\over 2F^2}\right)2\left(\Sl_{r=0}^MP_{A_{2r+1}}\right)\cdot p+2\left(\Sl_{r=0}^MP_{A_{2r+1}}\right)\cdot(p+q)S^{(0)}_{2,3}\Biggr]\prod\limits_{j=1}^{2M+1}J\left(A_{j}\right) \nn
&+&\Sl_{\substack{\{4,\dots,2n\}\\\to\{A_1,\dots,A_{2M+1}\}}}\left({-{1\over P^2_{4,2n}}}\right)\left(-{1\over 2F^2}\right)^M\Biggl[\left(\Sl_{r=0}^MP_{A_{2r+1}}\right)^2\left(S^{(1)}_{2,3}J(A_1)\right)\prod\limits_{s=2}^{2M+1}J(A_s)\Biggr].
\eea
Considering the explicit expression of $S^{(0)}_{2,3}$ in \eqref{Eq:On-shell-Leading-Double-Soft}, we apply the property \eqref{property1} to the first two terms in \eqref{EqGenAdjSubLeading} as well as the expression inside the brackets of the third term in \eqref{EqGenAdjSubLeading}. Then the first three terms together with the angular momentum part isolated from $S^{(1)}_{2,3}$ in the last term give
\bea
\left(-{1\over 2F^2}\right){q_{\mu}p_{\nu}{\cal J}_4^{\mu\nu}\over k_4\cdot(p+q)}J(4,\dots,2n),
\eea
 where we used the explicit Berends-Giele recursion of $J(4,\dots,n)$.
In the end, the coefficient of $\tau^1$ in \eqref{EqGenAdjSubLeading} is given by
 \bea
S^{(1)}_{2,3}J(4,\dots,2n),
 \eea
 which agrees with our expectation of the subleading order double-soft behavior of $J(\W 2,\W 3,4,\dots,2n)$ as shown in \eqref{Off-shell-Leading-Subleading-Double-Soft0}.
\end{itemize}

\subsubsection{Both soft particles are nonadjacent to the off-shell line}\label{subsubsect-nonadj}
%
\begin{figure}
  \centering
 \includegraphics[width=0.3\textwidth]{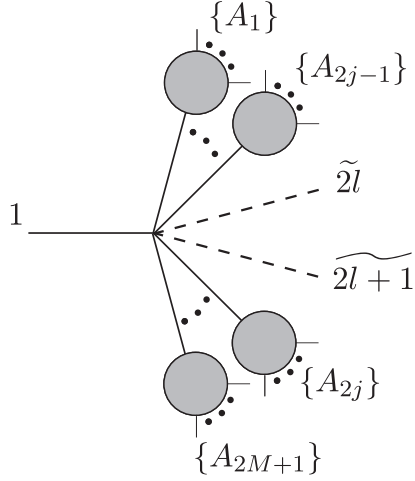}
 \caption{Type-1: diagrams with the soft particles ${2l}$ and ${2l+1}$  attached to the off-shell line directly} \label{FigGenNonAdjType1}
\end{figure}
\begin{figure}
  \centering
 \includegraphics[width=0.7\textwidth]{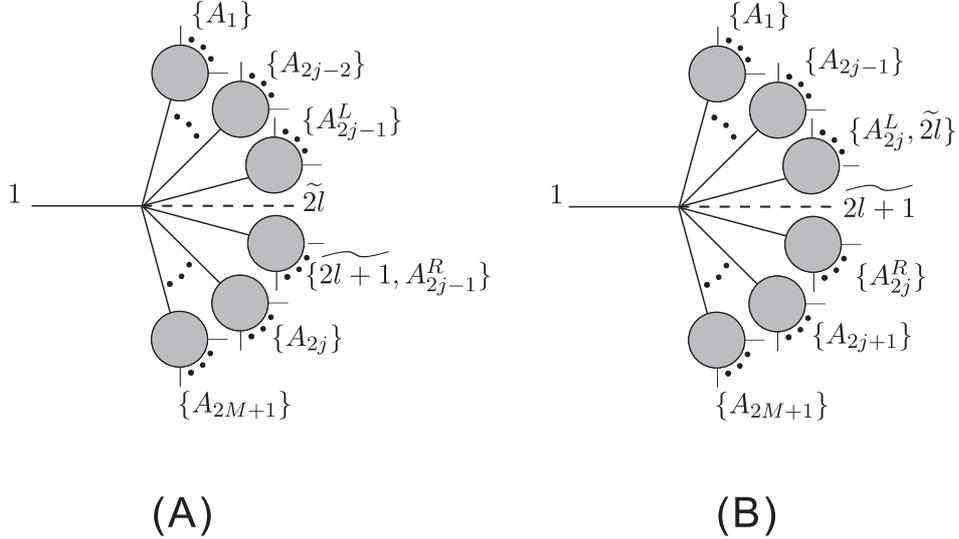}
 \caption{Type-2: diagrams with the soft particle ${2l}$ (or ${2l+1}$) attached to the off-shell line and the other one ${2l+1}$ (or
$ {2l}$) in a subcurrent} \label{FigGenNonAdjType2}
\end{figure}
\begin{figure}
  \centering
 \includegraphics[width=0.8\textwidth]
 {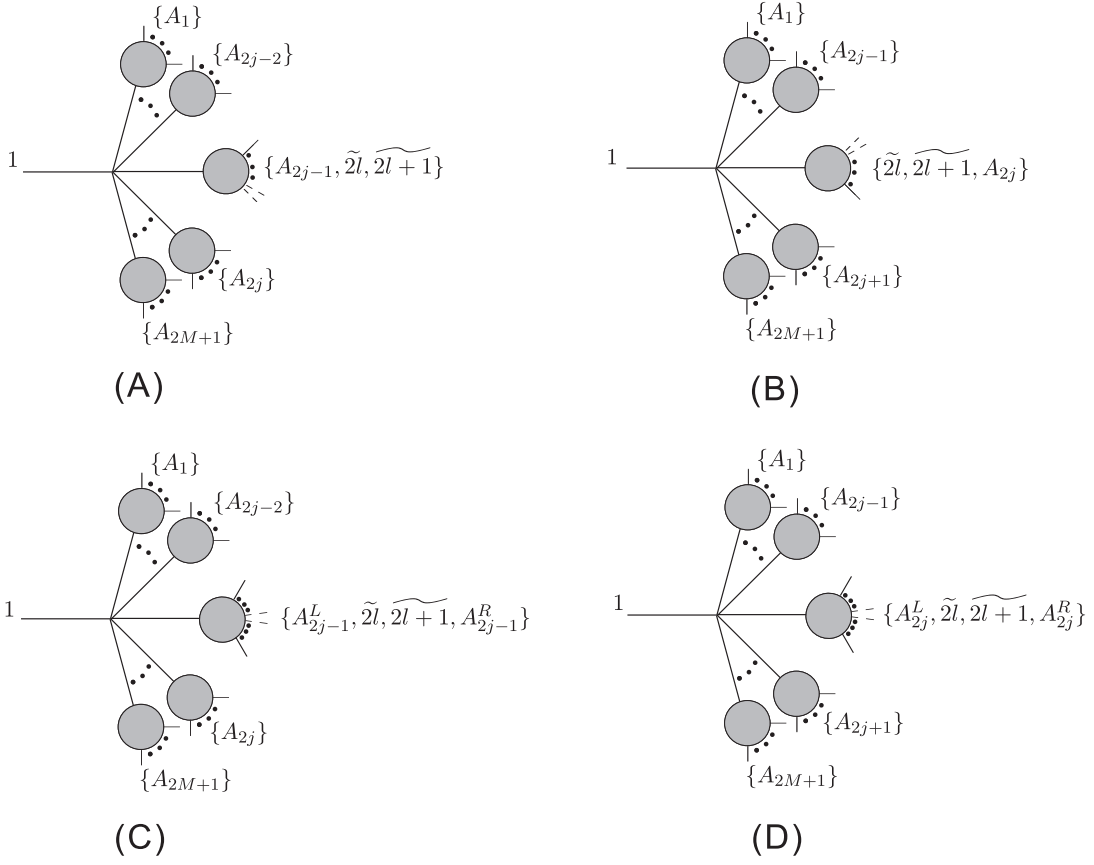}
 \caption{Type-3: diagrams with both soft particles ${2l}$ and ${2l+1}$ in a same subcurrent} \label{FigGenNonAdjType3}
\end{figure}
We have proved the double soft behavior of $J(\W 2,\W 3, 4, \dots, n)$, now let us study the behavior of $J(2,\dots,2l-1,\W {2l},\W {2l+1},2l+2,\dots,2n)$ $(1<l<n)$ with $2l$ and $2l+1$ soft.
Similar with what we have done in the boundary case with the soft particle $2$ adjacent to the off-shell line, we again classify the contributions in the Berends-Giele recursion into three types:
\begin{itemize}
\item \textit{{type-1}} diagrams with the soft particles ${2l}$ and ${2l+1}$ attached to the off-shell line directly, as shown in \figref{FigGenNonAdjType1},
\item  \textit{{type-2}} diagrams with the soft particle ${2l}$ (or ${2l+1}$) attached to the off-shell line and the other one ${2l+1}$ (or
$ {2l}$) in a subcurrent, as shown in \figref{FigGenNonAdjType2},
\item \textit{{type-3}} diagrams with both soft particles ${2l}$ and ${2l+1}$ in a same subcurrent, as  shown in \figref{FigGenNonAdjType3}.
\end{itemize}
Then the current is given by
\bea
&&J(2,\dots,2l-1,\W{2l},\W{2l+1},2l+2,\dots,2n)\Label{EqGenNonAdj} \\
&=&{i\over P^2_{2,2n}(k_{2l}=\tau p,k_{2l+1}=\tau q)}i\left[ T^{\text{N}}_{\text{Type-1}}(\tau)+T^{\text{N}}_{\text{Type-2}}(\tau)+T^{\text{N}}_{\text{Type-3}}(\tau)\right],\nonumber
\eea
where $T^{\text{N}}_{\text{Type-1}}(\tau)$, $T^{\text{N}}_{\text{Type-2}}(\tau)$ and $T^{\text{N}}_{\text{Type-3}}(\tau)$ denote the contributions from the above 3 types respectively.
The type-1 diagrams are typically expressed by \figref{FigGenNonAdjType1}, the type-2 diagrams are classified into (A) and (B) (see \figref{FigGenNonAdjType2}), while the type-3 diagrams are further classified into four types (A), (B), (C) and (D), as shown in \figref{FigGenNonAdjType3}.
Using the inductive assumptions for single and double soft behaviors of lower-point subcurrents, we express all contributions of figures \ref{FigGenNonAdjType1}, \ref{FigGenNonAdjType2} and \ref{FigGenNonAdjType3} explicitly and collect the results in \appref{app:appendix-nonadj}. The propagator of the off-shell line is expanded with respect to $\tau$ as
\bea
{1\over P^2_{2,2n}(k_{2l}=\tau p,k_{2l+1}=\tau q)}=\tau^0\left({1\over P^2_{2,2n}}\right)^{(0)}+\tau^1\left({1\over P^2_{2,2n}}\right)^{(1)}+\mathcal{O}(\tau^2),\Label{PropagatorGenNonAdj}
\eea
where the coefficients of $\tau^0$ and $\tau^1$ are
\bea
\left({1\over P^2_{2,2n}}\right)^{(0)}={1\over \left(P_{2,2l-1}+P_{2l+2,2n}\right)^2}\,, \qquad \left({1\over P^2_{2,2n}}\right)^{(1)}=-{2\left(P_{2,2l-1}+P_{2l+2,2n}\right)\cdot (p+q)\over \left[\left(P_{2,2l-1}+P_{2l+2,2n}\right)^2\right]^2}. \Label{PropagatorGenNonAdj1}
\eea
Inserting \eqref{PropagatorGenNonAdj} as well as $T^{\text{N}}_{\text{Type-1}}(\tau)$, $T^{\text{N}}_{\text{Type-2}}(\tau)$ and $T^{\text{N}}_{\text{Type-3}}(\tau)$ into \eqref{EqGenNonAdj}, we obtain the leading and subleading double soft behaviors as follows.
\begin{table}\centering
{\begin{tabular}{c|c|c|c}
  \hline
   &${i\over P^2_{2,2n}(\tau)}iT^{\text{N}}_{\text{Type-1}}(\tau)$ & ${i\over P^2_{2,2n}(\tau)}iT^{\text{N}}_{\text{Type-2}}(\tau)$ & ${i\over P^2_{2,2n}(\tau)}iT^{\text{N}}_{\text{Type-3}}(\tau)$ \\ \hline\hline
  $S^{(0)}J(2,\dots,\W{2l},\W{2l+1},\dots,2n)$ & $\substack{i\left({1\over P^2_{2,2n}}\right)^{(0)}iT^{\text{N}(0)}_{\text{Type-1}}}$  & $0$ & $\substack{i\left({1\over P^2_{2,2n}}\right)^{(0)}i \Bigl[T^{\text{N}(0)}_{\text{Type-3(A)}}+T^{\text{N}(0)}_{\text{Type-3(B)}}\\~~~~~~~~~~~~~~+T^{\text{N}(0)}_{\text{Type-3(C)}}+T^{\text{N}(0)}_{\text{Type-3(D)}}\Bigr]}$   \\ \hline
  $S_A^{(1)}J(2,\dots,\W{2l},\W{2l+1},\dots,2n)$ & 0  & 0  & $\substack{i\left({1\over P^2_{2,2n}}\right)^{(0)}i \Bigl[T^{\text{N}(1)}_{\text{Type-3(A2)}}+T^{\text{N}(1)}_{\text{Type-3(B1)}}\\~~~~~~~~~~~~~~~+T^{\text{N}(1)}_{\text{Type-3(C2)}}+T^{\text{N}(1)}_{\text{Type-3(D1)}}\Bigr]}$\\ \hline
  $S_B^{(1)}J(2,\dots,\W{2l},\W{2l+1},\dots,2n)$  & $
\substack{~~i\left({1\over P^2_{2,2n}}\right)^{(0)}iT^{\text{N}(1)}_{\text{Type-1}}\\+i\left({1\over P^2_{2,2n}}\right)^{(1)}iT^{\text{N}(0)}_{\text{Type-1}}}$   & 0  & $\substack{i\left({1\over P^2_{2,2n}}\right)^{(0)}i \Bigl[T^{\text{N}(1)}_{\text{Type-3(A1)}}+T^{\text{N}(1)}_{\text{Type-3(A3)}}\\~~~~~~~~~~~~~~+T^{\text{N}(1)}_{\text{Type-3(B2)}}+T^{\text{N}(1)}_{\text{Type-3(C1)}}\\~~~~~~~~~~~~~~~~+T^{\text{N}(1)}_{\text{Type-3(C3)}}+T^{\text{N}(1)}_{\text{Type-3(D2)}}\Bigr]\\
+i\left({1\over P^2_{2,2n}}\right)^{(1)}i\Bigl[T^{\text{N}(0)}_{\text{Type-3(A)}}+T^{\text{N}(0)}_{\text{Type-3(B)}}\\~~~~~~~~~~~~~~~~~+T^{\text{N}(0)}_{\text{Type-3(C)}}+T^{\text{N}(0)}_{\text{Type-3(D)}}\Bigr]}$\\ \hline
  $
     \begin{array}{c}
       \left({1\over 2F^2}\right)J(2,\dots,2l-1) \\
       \times J^{(1)}(\W{2l+1},2l+2,\dots,2n) \\
     \end{array}
$  &  0 & $\substack{i\left({1\over P^2_{2,2n}}\right)^{(0)}i\Bigl[T^{\text{N}(1)}_{\text{Type-2(A)}}\\+T^{\text{N}(1)}_{\text{Type-2(B)}}\Bigr]}$ & $\substack{\left({i\over P^2_{2,2n}}\right)^{(0)}i \Bigl[T^{\text{N}(1)}_{\text{Type-3(C4)}}+T^{\text{N}(1)}_{\text{Type-3(D3)}}\Bigr]}$  \\ \hline
\end{tabular}}\caption{The three types of contributions to the leading and subleading order double soft behaviors.}\Label{TableNonAdj}
\end{table}

\begin{itemize}
\item \textit{Leading order}

The leading order of $J(2,\dots, \W{2l},\W{2l+1},\dots,2n)$ is given by summing the second row of \tabref{TableNonAdj}.
The total contribution of the expression at row $2$, column $2$ and the first two terms at row $2$, column $4$ provides the sum of diagrams with particles $2l-1$, $2l+2$ in different subcurrents (i.e., \figref{FigGenNonAdjType1} and the diagrams (A), (B) in \figref{FigGenNonAdjType3})  in the Berends-Giele recursion of $J(2,\dots,2l-1,2l+2,\dots,2n)$. The coefficient of this term is
\bea
\left(-{1\over 2F^2}\right)+\left({1\over 2F^2}\right){k_{2l-1}\cdot q\over k_{2l-1}(p+q)}+\left({1\over 2F^2}\right){k_{2l+2}\cdot p\over k_{2l+2}(q+p)}=S_{2l,2l+1}^{(0)}.
\eea
The last two terms in row $2$ column $4$ produce the sum of diagrams with $2l-1$ and $2l+2$ in a same subcurrent in the expression of $J(2,\dots,2l-1,2l+2,\dots,2n)$. The accompanied coefficient is also $S_{2l,2l+1}^{(0)}$.
All together, the second row in \tabref{TableNonAdj} contributes an $S_{2l,2l+1}^{(0)}J(2,\dots,2l-1,2l+1,\dots,2n)$ which agrees with our expectation of the leading order term.

\item \textit{Subleading order}

The subleading order is obtained from nonzero terms from row $3$ to row $5$ in \tabref{TableNonAdj}.
\begin{enumerate}
\item
$S_A^{(1)}J(2,\dots,2l-1,2l+2,\dots,2n)$, where $S_{A}^{(1)}$ is defined by
 the first term of \eqref{Eq:On-shell-Subleading-Double-Soft}, only gains contribution from row $3$, column $4$ in \tabref{TableNonAdj}.
The first two terms there present all the diagrams of $J(2,\dots,2l-1,2l+1,\dots,2n)$ with particles $2l-1$, $2l+2$ in different subcurrents. The coefficient is given by
\bea
\left(-{1\over 2F^2}\right)(p\cdot q){k_{2l-1}\cdot q\over (k_{2l-1}\cdot (p+q))^2}+\left(-{1\over 2F^2}\right)(p\cdot q){k_{2l+2}\cdot q\over (k_{2l+2}\cdot (p+q))^2}=S_A^{(1)}.
\eea
The last two terms there gives the sum of all diagrams of $J(2,\dots,2l-1,2l+1,\dots,2n)$ with $2l-1$, $2l+2$ in a same subcurrent, and the prefactor is also $S_A^{(1)}$.
Thus the row $3$ precisely gives $S_A^{(1)}\, J(2,\dots,2l-1,2l+2,\dots,2n)$ as shown at row $3$, column $1$.

\item
$S_B^{(1)}J(2,\dots,2l-1,2l+2,\dots,2n)$, where $S_B^{(1)}$ is defined by the term containing angular momenta in equation \eqref{Eq:On-shell-Subleading-Double-Soft}, gains contribution from row $4$ in \tabref{TableNonAdj}. To check this result, we apply the property \eqref{property1} repeatedly and then find that
\bea
&&i\left({1\over P^2_{2,2n}}\right)^{(1)}\left[iT^{\text{N}(0)}_{\text{Type-1}}+i\left(T^{\text{N}(0)}_{\text{Type-3(A)}}+T^{\text{N}(0)}_{\text{Type-3(B)}}+T^{\text{N}(0)}_{\text{Type-3(C)}}+T^{\text{N}(0)}_{\text{Type-3(D)}}\right)\right]\Label{SB1-1}\nn
&=& {1\over (P_{2,2l-1}+P_{2l+2,2n})^2}\left[-S_B^{(1)}(P_{2,2l-1}+P_{2l+2,2n})^2\right]J(2,\dots,2l-1,2l+2,\dots,2n).\eea
Considering that $i\left({1\over P^2_{2,2n}}\right)^{(0)}i\left[T^{\text{N}(1)}_{\text{Type-1}}+ T^{\text{N}(1)}_{\text{Type-3(A1)}}+T^{\text{N}(1)}_{\text{Type-3(C1)}}\right]$ contains $S^{(1)}_B$ acting on the vertices, while $i\left({1\over P^2_{2,2n}}\right)^{(0)}i\Bigl[T^{\text{N}(1)}_{\text{Type-3(A3)}}+T^{\text{N}(1)}_{\text{Type-3(B2)}}+ T^{\text{N}(1)}_{\text{Type-3(C3)}}+T^{\text{N}(1)}_{\text{Type-3(D2)}}\Bigr]$ contains $S^{(1)}_B$ acting on different subcurrents, one obtains
\bea
&&i\left({1\over P^2_{2,2n}}\right)^{(0)}iT^{\text{N}(1)}_{\text{Type-1}}+i\left({1\over P^2_{2,2n}}\right)^{(0)}i\Bigl[T^{\text{N}(1)}_{\text{Type-3(A1)}}+T^{\text{N}(1)}_{\text{Type-3(A3)}}+T^{\text{N}(1)}_{\text{Type-3(B2)}}\nn
&&~~~~~~~~~~~~~~~~~~~~~~~~~~~~~~~~~~+ T^{\text{N}(1)}_{\text{Type-3(C1)}}+ T^{\text{N}(1)}_{\text{Type-3(C3)}}+T^{\text{N}(1)}_{\text{Type-3(D2)}}\Bigr]\nn
&=&{1\over (P_{2,2l-1}+P_{2l+2,2n})^2}\left[S_B^{(1)}(P_{2,2l-1}+P_{2l+2,2n})^2J(2,\dots,2l-1,2l+2,\dots,2n)\right].\Label{SB1-2}
\eea
Finally, the sum of the equations \eqref{SB1-1} and \eqref{SB1-2} which come from  row $4$ of  \tabref{TableNonAdj} gives $S_B^{(1)}J(2,\dots,2l-1,2l+2,\dots,2n)$.

\item
$\left(1\over 2F^2\right)J(2,\dots,2l-1)J^{(1)}(\widetilde{2l+1},2l+2,\dots,2n)$ is obtained from the row $5$ in \tabref{TableNonAdj}.
Cancelation happens between ${\text{Type-3(C4)}}$ from the diagram (C) in \figref{FigGenNonAdjType3} and ${\text{Type-2(A)}}$ from the diagram (A) in \figref{FigGenNonAdjType2}, and between ${\text{Type-3(D3)}}$ and ${\text{Type-2(B)}}$ as well.
The only term which cannot be canceled is the one with the off-shell line connected to a four-point vertex in the diagram (A) of figure \figref{FigGenNonAdjType2}. All in all, we have
\bea
T^{\text{N}(1)}_{\text{Type-3(C4)}}+T^{\text{N}(1)}_{\text{Type-2(A)}}&=&\left(1\over 2F^2\right)J(2,\dots,2l-1)J(\widetilde{2l+1},2l+2,\dots,2n),\nn
T^{\text{N}(1)}_{\text{Type-3(D3)}}+T^{\text{N}(1)}_{\text{Type-2(B)}}&=&0.
\eea
             \end{enumerate}

\end{itemize}
%

\section{Behaviors of amplitudes with one soft particle and two adjacent soft particles}\label{sec:on-shell-limit}
In  \secref{sec:offshell-single-soft} and \secref{sec:offshell-double-soft}, how the currents present with  single and double soft pions emitted has been elaborated. Now let us study the single and double soft behaviors of on-shell amplitudes, in other words, on-shell limits of off-shell currents.


When we consider the single and double soft behaviors of amplitudes, all external particles should be on-shell, i.e., $k^2_i=0$. Hence, we should first impose the on-shell condition of the off-shell line and then follow  similar discussions in sections \ref{sec:offshell-single-soft} and \ref{sec:offshell-double-soft}.
One may expect to obtain the behaviors in another way: by directly multiplying a $P_{2,2n}^2$ to the already known off-shell behaviors (see \eqref{Eq:Off-shell-Leading-Single-Soft}, \eqref{Off-shell-Leading-Subleading-Double-Soft0} and \eqref{Off-shell-Leading-Subleading-Double-Soft1}) and taking the on-shell limit $P_{2,2n}^2\rightarrow0$. However, we have to be careful with the order of the on-shell limit and the soft limit. For all diagrams in the current $J(2,\dots,\W i,\dots, 2n)$ $(2\leq i\leq 2n)$ with one soft particle and the current $J(2,\dots,\W i,\W{i+1},\dots,2n)$ $(2<i<2n-1)$ with two soft particles, the on-shell limit and the soft limit can be exchanged. Thus there is no ambiguity to obtain the behaviors of amplitudes by taking the on-shell limits of the already known soft behaviors of these currents, as shown in the introduction.

The more subtle case is the on-shell limit of the boundary current with one of the two soft particles adjacent to the off-shell line. As stated in the introduction,
in this case, the momentum of the particle adjacent to a soft one is chosen to be dependent on others. Among all diagrams of the boundary current, e.g., $J(\W 2,\W 3, 4,\dots, 2n)$, the diagram (B) in \figref{FigGenAdjType1} is special since it gives a different result when we exchange the order of soft limit and on-shell limit.

If we take the double soft limit of the diagram (B) in \figref{FigGenAdjType1} first, both the four-point vertex and the denominator of the off-shell propagator in $J(4,\dots, 2n)$ are nonzero because the on-shell condition $k_1^2=P^2_{2,2n}=0$ was not taken into account at this step. However, as claimed, when we consider the double soft behavior of amplitudes, $k_1^2=0$ should be imposed first. The four-point vertex and the off-shell propagator of $J(4,\dots, 2n)$ in the diagram (B) of \figref{FigGenAdjType1} with $k_1^2=0$ should be zero. Thus this case is adjusted as
\bea
&&\left({1\over 2F^2}\right){(\tau p+P_{4,2n})^2\over P^2_{4,2n}}\left[-P^2_{4,2n}J(4,\dots,2n)\right]\nn
&=&\left({1\over 2F^2}\right){(\tau q+k_1)^2\over (\tau p+\tau q+k_1)^2}A(4,\dots,2n)\nn
&=&\left({1\over 2F^2}\right)\left[{k_1\cdot q\over k_1\cdot (p+q)}-\tau (k_1\cdot q){p\cdot q \over \left(k_1\cdot (p+q)\right)^2}\right]A(4,\dots, 2n)+ \mathcal{O}(\tau^2),
\eea
where the momentum conservation and on-shell condition $k_1^2=P^2_{2,2n}=0$ have been considered. After this adjustment of diagram (B) in \figref{FigGenAdjType1}, we recall that $k_1$ is chosen as the dependent momentum and does not appear in the explicit amplitude expressions, thus, the differential operator with respect to $k_1$ in the angular momentum does not take effect. Finally, the double soft behavior of on-shell amplitude is precisely given by \eqref{Eq:On-shell-Double-Soft}.

\section{Double soft behavior of amplitudes with two nonadjacent soft particles}\label{sec:NonAdjSoft}
Having the leading and subleading behaviors of amplitudes (currents) with two adjacent soft factors, we now turn to the behaviors with two nonadjacent soft particles. While considering $J(2,\dots,\W i,\dots,\W j,\dots,2n)$, we have to encounter more cases according to different positions of soft particles $\W i$ and $\W j$. 
Particularly, we have following cases
\begin{itemize}
\item both soft particles $i$, $j$ are adjacent to the off-shell line
\item only one of soft particles $i$, $j$ is adjacent to the off-shell line
\item none of the soft particles is adjacent to the off-shell line
\end{itemize}
and either one of $i$, $j$ can be chosen as even or odd number. 
We claim that the soft behavior for off-shell current with two nonadjacent soft particles can be calculated out by Berends-Giele recursion. Nevertheless, to avoid repeated and tedious discussions, we will achieve the results via another approach, i.e., the KK relation \cite{Kleiss:1988ne} in NLSM \cite{Chen:2013fya,Chen:2014dfa}, 
\bea
A\left(1,2,\dots,i-1,i,i+1,\dots,2n\right)=\Sl_{\alpha\in OP\left(\{2,\dots,i-1\}\bigcup\{i+1,\dots,2n\}^T\right)}(-1)^{2n-i}A\left(1,\{\alpha\},i\right),\Label{KK-relation}
\eea
where the sum goes over all possible permutations $\alpha$ with keeping the relative orders  of elements in the set $\{2,\dots,i-1\}$ and reversing the relative orders of elements in the set $\{i+1,\dots,2n\}$. 

We focus on the on-shell amplitudes $A(1,\dots,\W i,\dots,\W j,\dots,2n)$ for convenience. According to different positions of $\W i$ and $\W j$, we reclassify the amplitudes into two categories.
\begin{itemize}
\item Amplitudes in which $\W i$ and $\W j$ do not share any adjacent particle.
\item Amplitudes in which $\W i$ and $\W j$ share one common adjacent particle.
\end{itemize}
Apparently,  the off-shell current with both soft particles adjacent to the off-shell line is just the off-shell version of the second category by taking the common particle as the off-shell leg.
Since partial amplitudes have cyclic symmetry, without loss of generality we can choose $1$ and $i$ (with $k_1=\tau p$, $k_i=\tau q$) as the two soft particles. In the the remaining part of this section, we will prove following double soft behaviors
\bea
A(\W 1, 2,\dots,i-1,\W i, i+1,\dots,2n)=0+\mathcal{O}(\tau^2),\qquad(3<i<2n-1) \Label{NonAdj-1}
\eea
\bea
A(\W 1, 2,\W 3,4,\dots,2n)=\Biggl\{
                             \begin{array}{cc}
                               0+\mathcal{O}(\tau^2) & (n=2) \\
                              \tau^1\left({1\over 2F^2}\right){p\cdot q\over k_2\cdot(p+q)}A(2,4,\dots,2n)+\mathcal{O}(\tau^2) & (n>2) \\
                             \end{array}.\Label{NonAdj-2}
\eea
All amplitudes with two nonadjacent soft particles can be obtained  by considering cyclic symmetry. 
Although the following proof is done by KK relation, we have alternatively verified these behaviors using Berends-Giele recursion. Since KK relation in nonlinear sigma model has already been extended to off-shell currents \cite{Chen:2014dfa}, one can also have a similar discussion for off-shell currents.

\subsection{Two soft particles do not share any common adjacent particle}
By the KK relation \eqref{KK-relation}, amplitude $A(\W 1, 2\dots i-1,\W i, i+1,\dots, 2n)$ where the two soft particles $\W 1$ and $\W i$ with momenta $k_1=\tau p$ and $k_i=\tau q$ do not share any adjacent particle can be rewritten as
\bea
A\left(1,2,\dots,i-1,i,i+1,\dots,2n\right)=\Sl_{\alpha\in OP\left(\{2,\dots,i-1\}\bigcup\{i+1,\dots,2n\}^T\right)}(-1)^{2n-i}A\left(1,\{\alpha\},i\right),\Label{KK-NonAdj1}
\eea
where $3<i<2n-1$. The right hand side of \eqref{KK-NonAdj1} is a combination of amplitudes with two adjacent soft particles $1$ and $i$.
The element $\alpha_1$ can be chosen as either $2$ or $2n$, while the element $\alpha_{2n-2}$ can be chosen as either $i-1$ or $i+1$.
Thus one can classify the amplitudes on the right hand side of \eqref{KK-NonAdj1} according to different choices of $(\alpha_1,\alpha_{2n-2})$ as following
\begin{itemize}
\item $\emph{(2,i-1)}$ This type of amplitudes has a general form $A(\W 1,2,\dots,i-1,\W i)$ and the total contribution of this type gives
\bea
I_1=\Sl_{\beta\in OP\left(\{3,\dots,i-2\}\bigcup\{i+1,\dots,2n\}^T\right)}(-1)^{2n-i}A\left(\W 1,2,\{\beta\},i-1,\W i\right).
\eea
Now, $A(\W 1,2,\dots,i-1,\W i)$ satisfy the two adjacent soft particles behaviors \eqref{Eq:On-shell-Double-Soft}, thus we arrive
\bea
I_1&=&\Sl_{\beta\in OP\left(\{3,\dots,i-2\}\bigcup\{i+1,\dots,2n\}^T\right)}(-1)^{2n-i}\left[\tau^0\mathbb{S}^{(0)}_{i,1}(k_{i-1},k_{ 2})+\tau^1\mathbb{S}^{(1)}_{i,1}(k_{i-1},k_{2})\right]A\left(2,\{\beta\},i-1\right)\nn
&&+\mathcal{O}(\tau^2),
\eea
where the double soft factors $\mathbb{S}^{(0)}_{i,1}(k_{i-1},k_{2})$ and $\mathbb{S}^{(1)}_{i,1}(k_{i-1},k_{ 2})$ are defined by \eqref{Eq:On-shell-Leading-Double-Soft} and \eqref{Eq:On-shell-Subleading-Double-Soft} and we rewrite them as functions of the momenta of those adjacent hard particles explicitly. After exchanging the summation and the soft factors, we obtain
\bea
I_1&=&(-1)^{2n-i}\left[\tau^0\mathbb{S}^{(0)}_{i,1}(k_{ i-1},k_{2})+\tau^1\mathbb{S}^{(1)}_{ i,1}(k_{i-1},k_{ 2})\right]\left[\Sl_{\beta\in OP\left(\{3,\dots,i-2\}\bigcup\{i+1,\dots,2n\}^T\right)}A\left(2,\{\beta\},i-1\right)\right]\nn
&&+\mathcal{O}(\tau^2),
\eea
The sum in the last brackets is nothing but the KK relation expression of $A(2,3,\dots,i-2,i-1,i+1,\dots,2n)$ up to a factor $(-1)^{2n-i}$, thus
\bea
I_1=\left[\tau^0\mathbb{S}^{(0)}_{i,1}(k_{ i-1},k_{2})+\tau^1\mathbb{S}^{(1)}_{ i,1}(k_{i-1},k_{ 2})\right]A(2,3,\dots,i-2,i-1,i+1,\dots,2n)+\mathcal{O}(\tau^2).
\eea
\item $\emph{(2,i+1)}$ Following a similar discussion in the $(2,i-1)$ case, we apply the double soft behavior with two adjacent soft particles as well as KK relation. Then the contribution of this type is written as
\bea
I_2&=&\Sl_{\beta\in OP\left(\{3,\dots,i-1\}\bigcup\{i+2,\dots,2n\}^T\right)}(-1)^{2n-i}A\left(\W 1,2,\{\beta\},i+1,\W i\right)\\
&=&(-1)\left[\tau^0\mathbb{S}^{(0)}_{i,1}(k_{i+1},k_{2})+\tau^1\mathbb{S}^{(1)}_{ i,1}(k_{i+1},k_{2})\right]A(2,3,\dots,i-2,i-1,i+1,\dots,2n)+\mathcal{O}(\tau^2).\nonumber
\eea
\item $\emph{(2n,i-1)}$ This type contributes
\bea
I_3&=&\Sl_{\beta\in OP\left(\{2,3,\dots,i-2\}\bigcup\{i+1,\dots,2n-1\}^T\right)}(-1)^{2n-i}A\left(\W 1,2n,\{\beta\},i-1,\W i\right)\\
&=&(-1)\left[\tau^0\mathbb{S}^{(0)}_{i,1}(k_{ i-1},k_{2n})+\tau^1\mathbb{S}^{(1)}_{ i,1}(k_{i-1},k_{2n})\right]A(2,3,\dots,i-2,i-1,i+1,\dots,2n)+\mathcal{O}(\tau^2).\nonumber
\eea
\item $\emph{(2n,i+1)}$ This type contributes
\bea
I_4&=&\Sl_{\beta\in OP\left(\{2,3,\dots,i-1\}\bigcup\{i+2,\dots,2n-1\}^T\right)}(-1)^{2n-i}A\left(\W 1,2n,\{\beta\},i+1,\W i\right)\\
&=&\left[\tau^0\mathbb{S}^{(0)}_{ i,1}(k_{i+1},k_{ 2n})+\tau^1\mathbb{S}^{(1)}_{i,1}(k_{i+1},k_{2n})\right]A(2,3,\dots,i-2,i-1,i+1,\dots,2n)+\mathcal{O}(\tau^2).\nonumber
\eea
\end{itemize}
Summing over all contributions together and considering the explicit forms of double soft factors \eqref{Eq:On-shell-Leading-Double-Soft} and \eqref{Eq:On-shell-Subleading-Double-Soft}, we finally find fully cancellation among the leading and subleading factors. Then we have proven the behavior \eqref{NonAdj-1}.

\subsection{Two soft particles share one common adjacent particle }
If two soft particles share one common adjacent particle, for example, $A(\W 1, 2,\W 3, 4,\dots,2n)$, we should isolatedly consider the behavior because in this case $2$ can be either chosen as $\alpha_1$ or $\alpha_{2n-2}$. A very special case is the four-point amplitude $A(\W 1,2,\W 3,4)$, where the two soft particles share two adjacent particles, namely $2$ and $4$. In this case, one can obtain the behavior directly by Feynman rules
\bea
A(\W 1,2,\W 3,4)=\mathcal{O}(\tau^2).
\eea

If $n>2$, the two soft particles share one common adjacent particle. We apply KK relation to rewrite $A(\W 1, 2,\W 3, 4,\dots,2n)$ as
\bea
A(\W 1, 2,\W 3, 4,\dots,2n)=(-1)^{2n-3}\Sl_{\alpha\in OP(\{2\}\bigcup\{4,\dots,2n\}^T)}A(\W 1,\{\alpha\},\W 3).
\eea
In this case, $(\alpha_1,\alpha_{2n-2})$ can be chosen as $(2,4)$, $(2n,4)$ or $(2n,2)$. 
In each type of contributions, we can apply the double soft behavior \eqref{Eq:On-shell-Double-Soft} as well as KK relation \eqref{KK-relation}. Then the behavior of $A(\W 1, 2,\W 3, 4,\dots,2n)$ reads
\bea
A(\W 1,2,\W 3,4,\dots,2n)&=&\bigl[\tau^0\Bigl(\mathbb{S}^{(0)}_{3,1}(k_{4},k_{2})-\mathbb{S}^{(0)}_{3,1}(k_{4},k_{2n})+\mathbb{S}^{(0)}_{3,1}(k_{2n},k_{2})\Bigr)\nn
&&+\tau^1\Bigl(\mathbb{S}^{(1)}_{3,1}(k_{4},k_{2})-\mathbb{S}^{(1)}_{3,1}(k_{4},k_{2n})+\mathbb{S}^{(1)}_{3,1}(k_{2n},k_{2})\Bigr)\Bigr]A(2,4,\dots,2n).
\eea
Substituting the definitions of the soft factors \eqref{Eq:On-shell-Leading-Double-Soft} and \eqref{Eq:On-shell-Subleading-Double-Soft} into the above equation,
we find that all the leading order factors as well as the angular momentum parts in the subleading factors cancel out. The left term gives out the behavior \eqref{KK-NonAdj1}.

\section{Conclusion}\label{sec:conclusion}

In this work, we studied the single and double soft behaviors of tree-level currents in NLSM.
We proved the leading behavior \eqref{Eq:Off-shell-Leading-Single-Soft} of currents with a single soft particle.
Furthermore, we proposed and proved the leading and subleading behaviors of currents containing two adjacent soft particles (see \eqref{Off-shell-Leading-Subleading-Double-Soft0} and \eqref{Off-shell-Leading-Subleading-Double-Soft1}).
The soft behaviors of the on-shell limits of currents were shown to be the right behaviors \eqref{Eq:On-shell-Leading-Single-Soft} and \eqref{Eq:On-shell-Double-Soft}. {\red By KK relation, we derived all double nonadjacent soft behaviors \eqref{NonAdj-1} and \eqref{NonAdj-2} for on-shell amplitudes.}
%

{\red This work provided a systematic study of the leading single soft behavior and the first two leading order double soft behaviors of tree amplitudes in NLSM by Berends-Giele recursion (Feynman diagrams) and KK relation.}

There are some interesting problems related to our present results:
\begin{itemize}
\item
As mentioned in the introduction, the inspiration of this subleading double soft behavior in the effective theory comes from a) the study of global symmetry breaking from an amplitude point of view and b) the recent soft theorems research in different frameworks.
Actually, the leading order behavior can be studied by current algebra as well \cite{Weinberg:1996kr, Cheng:1985bj}. 
Concerning the subleading results derived in this paper, we are curious on several points: how to understand these results from the current algebra point of view? Do these nice structures indicate some hidden symmetry or deeper physics insight?
\item Although, the leading and subleading order double soft behaviors as well as the leading order single soft behavior of currents have been proved, the subleading order of a current with a single soft particle is still unknown. Thus it is worthy studying the subleading order single soft behavior.
\item The behaviors of currents with more soft particles will be complicated and deserves further study.
\item All discussions in this work are limited at tree level. Loop level extension is expected.
\end{itemize}

\subsection*{Acknowledgements}
We thank Congkao Wen for his useful discussion and feedback on the manuscript.
Y.D. would  like to acknowledge the supports from Erasmus Mundus Action 2, Project 9, the International Postdoctoral Exchange Fellowship Program of China (with Fudan University as the home university), the NSF of China Grant No. 11105118, China Postdoctoral Science Foundation No. 2013M530175 and the Fundamental Research Funds for the Central Universities of Fudan University No. 20520133169.
H.L. is supported by the ERC Advanced Grant no. 267985 (DaMeSyFla).

\appendix

\section{ Feynman rules and Berends-Giele recursion in NLSM}\label{app:Feynman-rules-BG}
We review Feynman rules and Berends-Giele recursion in NLSM.
Most of the notations agree with those in \cite{Kampf:2012fn,Kampf:2013vha}. This part has some overlaps with \cite{Chen:2013fya,Chen:2014dfa}.

\subsection{Feynman rules}

{~~~~~\emph {Lagrangian}}

The Lagrangian of $U(N)$ non-linear sigma model is
\bea
\mathcal{L}={F^2\over 4}\mathrm{Tr} (\partial_{\mu}U\partial^{\mu}U^{\dagger}),
\eea
where $F$ is a constant of one mass dimension. Using Caylay parametrization as in \cite{Kampf:2012fn,Kampf:2013vha}, the $U$ is defined by
\bea
U=1+2\Sl_{n=1}^{\infty}\left({1\over 2F}\phi\right)^n,~~~~\label{Cayley}
\eea
where $\phi=\sqrt{2}\phi^at^a$ and $t^a$ are generators of $U(N)$ Lie algebra.

{\emph {Color decomposition}}

The full tree amplitudes can be given by a trace form color decomposition\footnote{In \cite{Kampf:2012fn} and \cite{Kampf:2013vha}, this decomposition is mentioned as flavor decomposition.}
\bea
M(1^{a_1},\dots,n^{a_n})=\Sl_{\sigma\in S_{n-1}}\mathrm{Tr}(T^{a_{1}}T^{a_{\sigma_2}}\dots T^{a_{\sigma_n}})A(1,\sigma).\label{Trace form}
\eea
Since traces have cyclic symmetry, the partial amplitudes $A$ also satisfy cyclic symmetry
\bea
A(1,2,\dots,n)=A(n,1,\dots,n-1).\label{Cyclic symmetry}
\eea

\emph{Feynman rules for partial amplitudes}

Under Cayley parametrization \eqref{Cayley}, vertices in the Feynman rules for partial amplitudes are
\bea
V_{2n+1}=0,\qquad V_{2n+2}=\left(-{1\over 2F^2}\right)^n\left(\Sl_{i=0}^np_{2i+1}\right)^2=\left(-{1\over 2F^2}\right)^n\left(\Sl_{i=0}^np_{2i+2}\right)^2.\label{Feyn-rules}
\eea
Here, $p_j$ denotes the momentum of the leg $j$; momentum conservation has been considered.

\subsection{Berends-Giele recursion}
From the above Feynman rules, one can construct tree-level currents\footnote{In this paper, an $n$-point current is mentioned as the current with $n-1$ on-shell legs and one off-shell leg.} through Berends-Giele recursion
\bea
&&J(2,...,2n)\\
&=&\frac{i}{P_{2,2n}^2}\Sl_{m=2}^n\Sl_{\text{Divisions}}i V_{2m}(p_1=-P_{2,2n},P_{A_1},\cdots,P_{A_{2m-1}})\times\prod\limits_{k=1}^{2m-1} J(A_{k}),\label{B-G}\nonumber
\eea
where $p_1=-P_{2,n}=-(p_2+p_3+\dots+p_n)$ is the momentum of the off-shell leg $1$. In the second sum,  'Divisions' denote all possible divisions
of on-shell particles $\{2,\dots,2n\}\to \{A_1\},\dots,\{A_{2m-1}\}$ with odd number elements in each subset. For example, if $n=6$ and $m=2$, we have three divisions $\{2,3,4,5,6\}\to \{2,3,4\},\{5\},\{6\}$, $\{2,3,4,5,6\}\to \{2\},\{3,4,5\},\{6\}$ and $\{2,3,4,5,6\}\to \{2\},\{3\},\{4,5,6\}$. The starting point of this recursion is $J(2)=J(3)=\dots=J(n)=1$.

 Since there is at least one odd-point vertex for a current containing odd lines (including the off-shell line), according to the vanishing odd-point vertex as in \eqref{Feyn-rules}, we always have
\bea
J(2,\dots,2n+1)=0,
\eea
for $(2n+1)$-point amplitudes.
The currents with even points are nonzero in general and are built up by only odd numbers of even-point subcurrents.
Due to the vanishment of odd-point currents,
our discussion in this paper only concern even-point currents.

The current apparently have reflection symmetry
\bea
J(2,3,\dots,2n)=J(2n,\dots,3,2).\Label{eq:reflection}
\eea
\section{Computation details for nonadjacent case in section \ref{subsubsect-nonadj}}\label{app:appendix-nonadj}
In this section, we illustrate all computation details from diagrams in figures \ref{FigGenNonAdjType1}, \ref{FigGenNonAdjType2} and \ref{FigGenNonAdjType3}.

\textit{Type-1} The type-1 diagrams with two soft particles $i$, $i+1$ attached to the off-shell line are typically described by \figref{FigGenNonAdjType1}. When $\tau\to 0$, the sum of diagrams in this type is expanded as
    \bea
    T^{\text{N}}_{\text{Type-1}}(\tau)= \tau^0 T^{\text{N}(0)}_{\text{Type-1}}+\tau^1 T^{\text{N}(1)}_{\text{Type-1}},
    \eea
    where $ T^{\text{N}(0)}_{\text{Type-1}}$ and $T^{\text{N}(1)}_{\text{Type-1}}$ are given by
    \bea
    T^{\text{N}(0)}_{\text{Type-1}}&=&\Sl_{\substack{\{2,\dots,2l-1\}\to\{A_1\},\dots,\{A_{2j-1}\}\\\{2l+2,\dots,2n\}\to\{A_{2j}\},\dots,\{A_{2M+1}\}}}\left(-{1\over 2F^2}\right)^{M+1}\left(\Sl_{r=0}^{M+1}P_{2r-1}\right)^2\prod\limits_{s=1}^{2M+1} J(A_s),\\
    T^{\text{N}(1)}_{\text{Type-1}}&=&\Sl_{\substack{\{2,\dots,2l-1\}\to\{A_1\},\dots,\{A_{2j-1}\}\\\{2l+2,\dots,2n\}\to\{A_{2j}\},\dots,\{A_{2M+1}\}}}\left(-{1\over 2F^2}\right)^{M+1}2\left(\Sl_{r=0}^{M+1}P_{2r-1}\right)\cdot q \prod\limits_{s=1}^{2M+1} J(A_s),
    \eea
    where we have summed over all possible divisions $\{2,\dots,2l-1\}\to\{A_1\},\dots,\{A_{2j-1}\}$ and $\{2l+2,\dots,2n\}\to\{A_{2j}\},\dots,\{A_{2M+1}\}$ with elements of odd number in each set.

\textit{Type-2} Type-2 diagrams are shown by two diagrams in \figref{FigGenNonAdjType2}.
The diagram (A) of \figref{FigGenNonAdjType2} stands for those with ${2l}$ connected to the off-shell line and $2l+1$ in a nontrivial subcurrent, while the diagram (B) has ${2l}$ in a nontrivial subcurrent but ${2l+1}$ connected to the off-shell line. As proven before, the behavior of  currents  of the forms $J(\W{2l+1},\dots)$ and $J(\dots,\W{2l})$ have to vanish. The $\tau^0$ order term should be zero and the $\tau^1$ order term of the sum of all type-2 diagrams is
    \bea
    T^{\text{N}}_{\text{Type-2}}(\tau)&=&\tau^1\left[T^{{\text{N}}(1)}_{\text{Type-2(A)}}+T^{{\text{N}}(1)}_{\text{Type-2(B)}}\right],
    \eea
Here, $T^{{\text{N}}(1)}_{\text{Type-2(A)}}$ is the sum of $\tau^1$ coefficient of all possible (A) diagrams in \figref{FigGenNonAdjType2} and given by
\bea
T^{{\text{N}}(1)}_{\text{Type-2(A)}}&=&\Sl_{\substack{\{2,\dots,2l-1\}\to\{A_1\},\dots,\{A_{2j-2}\},\{A^L_{2j-1}\}\\\{2l+2,\dots,2n\}\to\{A^R_{2j-1}\},\{A_{2j}\},\dots,\{A_{2M+1}\}}}\left(-{1\over 2F^2}\right)^{M+1}\\
&&~~~~~~~~~~~~~~~~~~~~~~\times\left(\Sl_{r=0}^MP_{A_{2r+1}}\right)^2J(A^L_{2j-1})J(\W{2l+1},A^R_{2j-1})\prod\limits_{\substack{s=1\\s\neq 2j-1}}^{2M+1}J(A_s),\nonumber
\eea
where we have defined $P_{A_{2j-1}}\equiv P_{A^L_{2j-1}}+P_{A^R_{2j-1}}$. We emphasize that there should be even number of elements in $\{A^R_{2j-1}\}$ because $\{\W{2l+1},A^R_{2j-1}\}$ should contain odd number of elements.  The sum of all possible (B) diagrams in \figref{FigGenNonAdjType2}, $T^{(1)}_{\text{Type-2(B)}}(\W {2l},\W {2l+1})$, is expressed by
\bea
T^{{\text{N}}(1)}_{\text{Type-2(B)}}&=&\Sl_{\substack{\{2,\dots,2l-1\}\to\{A_1\},\dots,\{A_{2j-1}\},\{A^L_{2j}\}\\\{2l+2,\dots,2n\}\to\{A^R_{2j}\},\{A_{2j+1}\},\dots,\{A_{2M+1}\}}}\left(-{1\over 2F^2}\right)^{M+1}\\
&&~~~~~~~~~~~~~~~~~~~~~~\times\left(\Sl_{r=0}^MP_{A_{2r+1}}\right)^2J(A^L_{2j})J(\W{2l+1},A^R_{2j})\prod\limits_{\substack{s=1\\s\neq 2j}}^{2M+1}J(A_s),\nonumber
\eea
where $\{A^R_{2j}\}$ contains even number of elements.

\textit{Type-3} When two soft particles are in a same subcurrent, we have following four kinds of diagrams (see the four diagrams in \figref{FigGenNonAdjType3}).

\begin{itemize}
\item \textit{Diagrams containing a subcurrent with the form $J(\dots,\W{2l},\W{2l+1})$}

In this case, we substitute the double soft behavior of lower-point currents with ${2l+1}$ adjacent to the propagator. The sum of diagrams of this kind is written as
    \bea
     T^{\text{N}}_{\text{Type-3(A)}}(\tau)&=&\tau^0 T^{{\text{N}}(0)}_{\text{Type-3(A)}}+\tau^1\Bigl[T^{{\text{N}}(1)}_{\text{Type-3(A1)}}+T^{{\text{N}}(1)}_{\text{Type-3(A2)}}+T^{{\text{N}}(1)}_{\text{Type-3(A3)}}\Bigr],
    \eea
    where
    \bea
    T^{{\text{N}}(0)}_{\text{Type-3(A)}}&=&\Sl_{\substack{\{2,\dots,2l-1\}\to\{A_1\}\dots\{A_{2j-1}\}\\\{2l+2,\dots,2n\}\to\{A_{2j}\}\dots\{A_{2M+1}\}}}\left({1\over 2F^2}\right){k_{2l-1}\cdot q\over k_{2l-1}\cdot (p+q)}\\
    &&~~~~~~~~~~~~~~~~~~~~~~\times\left(-{1\over 2F^2}\right)^{M}\left(\Sl_{r=0}^{M}P_{A_{2r+1}}\right)^2\prod_{s=1}^{2M+1}J(A_s),\nn
    T^{{\text{N}}(1)}_{\text{Type-3(A1)}}&=&\Sl_{\substack{\{2,\dots,2l-1\}\to\{A_1\}\dots\{A_{2j-1}\}\\\{2l+2,\dots,2n\}\to\{A_{2j}\}\dots\{A_{2M+1}\}}}\left({1\over 2F^2}\right){k_{2l-1}\cdot q\over k_{2l-1}\cdot (p+q)}\\
    &&~~~~~~~~~~~~~~~~~~~~~~\times\left(-{1\over 2F^2}\right)^{M}2\left(\Sl_{r=0}^{M}P_{A_{2r+1}}\right)\cdot(p+q)\prod_{s=1}^{2M+1}J(A_s),\nn
    T^{{\text{N}}(1)}_{\text{Type-3(A2)}}&=&\Sl_{\substack{\{2,\dots,2l-1\}\to\{A_1\}\dots\{A_{2j-1}\}\\\{2l+2,\dots,2n\}\to\{A_{2j}\}\dots\{A_{2M+1}\}}}\left(-{1\over 2F^2}\right)(p\cdot q){k_{2l-1}\cdot q\over (k_{2l-1}\cdot (p+q))^2}\\
    &&~~~~~~~~~~~~~~~~~~~~~~\times\left(-{1\over 2F^2}\right)^{M}\left(\Sl_{r=0}^{M}P_{A_{2r+1}}\right)^2\prod_{s=1}^{2M+1}J(A_s),\nn
    T^{\text{N}(1)}_{\text{Type-3(A3)}}&=&\Sl_{\substack{\{2,\dots,2l-1\}\to\{A_1\}\dots\{A_{2j-1}\}\\\{2l+2,\dots,2n\}\to\{A_{2j}\}\dots\{A_{2M+1}\}}}\left(-{1\over 2F^2}\right){q_{\mu}p_{\nu}{\cal J}_{2l-1}^{\mu\nu}\over k_{2l-1}\cdot (q+p)}J(A_{2j-1})\\
    &&~~~~~~~~~~~~~~~~~~~~~~\times\left(-{1\over 2F^2}\right)^{M}\left(\Sl_{r=0}^{M}P_{A_{2r+1}}\right)^2\prod_{\substack{s=1\\s\neq 2j-1}}^{2M+1}J(A_s).\nonumber
    \eea

\item \textit{Diagrams containing a subcurrent with the form $J(\W{2l},\W{2l+1},\dots) $}

We apply the soft behavior to the lower-point subcurrent with $2l$ adjacent to the off-shell line. The contribution of this type is then given by
    \bea
     T^{\text{N}}_{\text{Type-3(B)}}(\tau)&=&\tau^0 T^{{\text{N}}(0)}_{\text{Type-3(B)}}+\tau^1\Bigl[T^{{\text{N}}(1)}_{\text{Type-3(B1)}}+T^{{\text{N}}(1)}_{\text{Type-3(B2)}}\Bigr],
    \eea
    where
    \bea
    T^{{\text{N}}(0)}_{\text{Type-3(B)}}&=&\Sl_{\substack{\{2,\dots,2l-1\}\to\{A_1\}\dots\{A_{2j-1}\}\\\{2l+2,\dots,2n\}\to\{A_{2j}\}\dots\{A_{2M+1}\}}}\left({1\over 2F^2}\right){k_{2l+2}\cdot p\over k_{2l+2}\cdot (q+p)}\\
    &&~~~~~~~~~~~~~~~~~~~~~~\times\left(-{1\over 2F^2}\right)^{M}\left(\Sl_{r=0}^{M}P_{A_{2r+1}}\right)^2\prod_{s=1}^{2M+1}J(A_s),\nn
    T^{{\text{N}}(1)}_{\text{Type-3(B1)}}&=&\Sl_{\substack{\{2,\dots,2l-1\}\to\{A_1\}\dots\{A_{2j-1}\}\\\{2l+2,\dots,2n\}\to\{A_{2j}\}\dots\{A_{2M+1}\}}}\left(-{1\over 2F^2}\right)(p\cdot q){k_{2l+2}\cdot p\over (k_{2l+2}\cdot (q+p))^2}\\
    &&~~~~~~~~~~~~~~~~~~~~~~\times\left(-{1\over 2F^2}\right)^{M}\left(\Sl_{r=0}^{M}P_{A_{2r+1}}\right)^2\prod_{s=1}^{2M+1}J(A_s),\nn
    T^{{\text{N}}(1)}_{\text{Type-3(B2)}}&=&\Sl_{\substack{\{2,\dots,2l-1\}\to\{A_1\}\dots\{A_{2j-1}\}\\\{2l+2,\dots,2n\}\to\{A_{2j}\}\dots\{A_{2M+1}\}}}\left(-{1\over 2F^2}\right){p_{\mu}q_{\nu}{\cal J}_{2l+2}^{\mu\nu}\over k_{2l+2}\cdot (p+q)}J(A_{2j})\\
    &&~~~~~~~~~~~~~~~~~~~~~~\times\left(-{1\over 2F^2}\right)^{M}\left(\Sl_{r=0}^{M}P_{A_{2r+1}}\right)^2\prod_{\substack{s=1\\s\neq 2j}}^{2M+1}J(A_s).\nonumber
    \eea

\item \textit{Diagrams containing a subcurrent of the form $J(\dots,\W{2l},\W{2l+1},\dots)$ with odd numbers of elements on the left of ${2l}$ and even number of elements on the right of ${2l+1}$}

The total contribution from diagrams of this kind is given by
    \bea
     T^{\text{N}}_{\text{Type-3(C)}}&=&\tau^0 T^{{\text{N}}(0)}_{\text{Type-3(C)}}+\tau^1\Bigl[T^{{\text{N}}(1)}_{\text{Type-3(C1)}}+T^{{\text{N}}(1)}_{\text{Type-3(C2)}}+T^{{\text{N}}(1)}_{\text{Type-3(C3)}}+T^{{\text{N}}(1)}_{\text{Type-3(C4)}}\Bigr],
    \eea
where
\bea
T^{{\text{N}}(0)}_{\text{Type-3(C)}}&=&\Sl_{\substack{\{2,\dots,2l-1\}\to\{A_1\},\dots,\{A_{2j-2}\},\{A^L_{2j-1}\}\\\{2l+2,\dots,2n\}\to\{A^R_{2j-1}\},\{A_{2j}\}\dots,\{A_{2M+1}\}}}\left(-{1\over 2F^2}\right){1\over 2}\left[{k_{2l+2}\cdot(q-p)\over k_{2l+2}\cdot(q+p)}+{k_{2l-1}\cdot(p-q)\over k_{2l-1}\cdot(p+q)}\right]\nn
&&~~~~~~~~~~~~~~~~~~~~~~\times\left(-{1\over 2F^2}\right)^M\left(\Sl_{r=0}^MP_{A_{2r+1}}\right)^2\prod\limits_{s=1}^{2M+1}J(A_s),\\
T^{{\text{N}}(1)}_{\text{Type-3(C1)}}&=&\Sl_{\substack{\{2,\dots,2l-1\}\to\{A_1\},\dots,\{A_{2j-2}\},\{A^L_{2j-1}\}\\\{2l+2,\dots,2n\}\to\{A^R_{2j-1}\},\{A_{2j}\}\dots,\{A_{2M+1}\}}}\left(-{1\over 2F^2}\right){1\over 2}\left[{k_{2l+2}\cdot(q-p)\over k_{2l+2}\cdot(q+p)}+{k_{2l-1}\cdot(p-q)\over k_{2l-1}\cdot(p+q)}\right]\nn
&&~~~~~~~~~~~~~~~~~~~~~~\times\left(-{1\over 2F^2}\right)^M2\left(\Sl_{r=0}^MP_{A_{2r+1}}\right)\cdot(p+q)\prod\limits_{s=1}^{2M+1}J(A_s),\\
T^{{\text{N}}(1)}_{\text{Type-3(C2)}}&=&\Sl_{\substack{\{2,\dots,2l-1\}\to\{A_1\},\dots,\{A_{2j-2}\},\{A^L_{2j-1}\}\\\{2l+2,\dots,2n\}\to\{A^R_{2j-1}\},\{A_{2j}\}\dots,\{A_{2M+1}\}}}\left(-{1\over 2F^2}\right)(p\cdot q)\left[{k_{2l+2}\cdot p\over (k_{2l+2}\cdot(q+p))^2}+{k_{2l-1}\cdot q\over (k_{2l-1}\cdot(p+q))^2}\right]\nn
&&~~~~~~~~~~~~~~~~~~~~~~\times\left(-{1\over 2F^2}\right)^M\left(\Sl_{r=0}^MP_{A_{2r+1}}\right)^2\prod\limits_{s=1}^{2M+1}J(A_s),\\
T^{{\text{N}}(1)}_{\text{Type-3(C3)}}&=&\Sl_{\substack{\{2,\dots,2l-1\}\to\{A_1\},\dots,\{A_{2j-2}\},\{A^L_{2j-1}\}\\\{2l+2,\dots,2n\}\to\{A^R_{2j-1}\},\{A_{2j}\}\dots,\{A_{2M+1}\}}}\left(-{1\over 2F^2}\right)\left[{q_{\mu}p_{\nu}{\cal J}_{2l+2}^{\mu\nu}\over k_{2l+2}\cdot(q+p)}+{p_{\mu}q_{\nu}{\cal J}^{\mu\nu}_{2l-1}\over k_{2l-1}\cdot(p+q)}\right]J(A_{2j-1})\nn
&&~~~~~~~~~~~~~~~~~~~~~~\times\left(-{1\over 2F^2}\right)^M\left(\Sl_{r=0}^MP_{A_{2r+1}}\right)^2\prod\limits_{\substack{s=1\\ s\neq 2j-1}}^{2M+1}J(A_s),\\
T^{{\text{N}}(1)}_{\text{Type-3(C4)}}&=&\Sl_{\substack{\{2,\dots,2l-1\}\to\{A_1\},\dots,\{A_{2j-2}\},\{A^L_{2j-1}\}\\\{2l+2,\dots,2n\}\to\{A^R_{2j-1}\},\{A_{2j}\}\dots,\{A_{2M+1}\}}}\left({1\over 2F^2}\right)J(A^L_{2j-1})J(\W{2l+1},A^R_{2j-1})\nn
&&~~~~~~~~~~~~~~~~~~~~~~\times\left(-{1\over 2F^2}\right)^M\left(\Sl_{r=0}^MP_{A_{2r+1}}\right)^2\prod\limits_{\substack{s=1\\ s\neq 2j-1}}^{2M+1}J(A_s),
\eea
where $P_{A_{2j-1}}=P^L_{A_{2j-1}}+P^R_{A_{2j-1}}$.

\item \textit{Diagrams containing a subcurrent with the form $J(\dots,\W{2l},\W{2l+1},\dots)$ with even numbers of elements on the left of ${2l}$ and odd numbers of elements on the right of $2l+1$}

The total contribution of diagrams of this kind is given by
    \bea
     T^{\text{N}}_{\text{Type-3(D)}}&=&\tau^0 T^{\text{N}(0)}_{\text{Type-3(D)}}+\tau^1\Bigl[T^{\text{N}(1)}_{\text{Type-3(D1)}}+T^{\text{N}(1)}_{\text{Type-3(D2)}}+T^{\text{N}(1)}_{\text{Type-3(D3)}}\Bigr],
    \eea
where
\bea
T^{\text{N}(0)}_{\text{Type-3(D)}}&=&\Sl_{\substack{\{2,\dots,2l-1\}\to\{A_1\},\dots,\{A_{2j-1}\},\{A^L_{2j}\}\\\{2l+2,\dots,2n\}\to\{A^R_{2j}\},\{A_{2j+1}\}\dots,\{A_{2M+1}\}}}\left(-{1\over 2F^2}\right){1\over 2}\left[{k_{2l+2}\cdot(q-p)\over k_{2l+2}\cdot(q+p)}+{k_{2l-1}\cdot(p-q)\over k_{2l-1}\cdot(p+q)}\right]\nn
&&~~~~~~~~~~~~~~~~~~~~~~\times\left(-{1\over 2F^2}\right)^M\left(\Sl_{r=0}^MP_{A_{2r+1}}\right)^2\prod\limits_{s=1}^{2M+1}J(A_s),\\
T^{\text{N}(1)}_{\text{Type-3(D1)}}&=&\Sl_{\substack{\{2,\dots,2l-1\}\to\{A_1\},\dots,\{A_{2j-1}\},\{A^L_{2j}\}\\\{2l+2,\dots,2n\}\to\{A^R_{2j}\},\{A_{2j+1}\}\dots,\{A_{2M+1}\}}}\left(-{1\over 2F^2}\right)(p\cdot q)\left[{k_{2l+2}\cdot p\over (k_{2l+2}\cdot(q+p))^2}+{k_{2l-1}\cdot q\over (k_{2l-1}\cdot(p+q))^2}\right]\nn
&&~~~~~~~~~~~~~~~~~~~~~~\times\left(-{1\over 2F^2}\right)^M\left(\Sl_{r=0}^MP_{A_{2r+1}}\right)^2\prod\limits_{s=1}^{2M+1}J(A_s),\\
T^{{\text{N}}(1)}_{\text{Type-3(D2)}}
&=&\Sl_{\substack{\{2,\dots,2l-1\}\to\{A_1\},\dots,\{A_{2j-1}\},\{A^L_{2j}\}\\\{2l+2,\dots,2n\}\to\{A^R_{2j}\},\{A_{2j+1}\}\dots,\{A_{2M+1}\}}}\left(-{1\over 2F^2}\right)\left[{q_{\mu}p_{\nu}{\cal J}_{2l+2}^{\mu\nu}\over k_{2l+2}\cdot(q+p)}+{p_{\mu}q_{\nu}{\cal J}^{\mu\nu}_{2l-1}\over k_{2l-1}\cdot(p+q)}\right]J(A_{2j})\nn
&&~~~~~~~~~~~~~~~~~~~~~~\times\left(-{1\over 2F^2}\right)^M\left(\Sl_{r=0}^MP_{A_{2r+1}}\right)^2\prod\limits_{\substack{s=1\\ s\neq 2j}}^{2M+1}J(A_s),\\
T^{{\text{N}}(1)}_{\text{Type-3(D3)}}&=&\Sl_{\substack{\{2,\dots,2l-1\}\to\{A_1\},\dots,\{A_{2j-1}\},\{A^L_{2j}\}\\\{2l+2,\dots,2n\}\to\{A^R_{2j}\},\{A_{2j+1}\}\dots,\{A_{2M+1}\}}}\left({1\over 2F^2}\right)J(A^L_{2j},\W{2l})J(\W{2l+1},A^R_{2j})\nn
&&~~~~~~~~~~~~~~~~~~~~~~\times\left(-{1\over 2F^2}\right)^M\left(\Sl_{r=0}^MP_{A_{2r+1}}\right)^2\prod\limits_{\substack{s=1\\ s\neq 2j}}^{2M+1}J(A_s).
\eea
\end{itemize}

\bibliographystyle{JHEP}
\bibliography{Refs}

\end{document}